%% file: main.tex
\documentclass[lettersize,journal]{IEEEtran}
\IEEEoverridecommandlockouts
\usepackage{cite}
\usepackage{amsmath,amssymb,amsfonts}
\usepackage{algorithmic}
\usepackage{graphicx}
\usepackage{textcomp}
\usepackage{xcolor}
\def\BibTeX{{\rm B\kern-.05em{\sc i\kern-.025em b}\kern-.08em
    T\kern-.1667em\lower.7ex\hbox{E}\kern-.125emX}}
\input{preamble}
\usepackage{balance}
\begin{document}

\title{Can Large Language Models Serve as Evaluators for Code Summarization?}

\author{
Yang Wu,
Yao Wan*,
Zhaoyang Chu,
Wenting Zhao,
Ye Liu,
Hongyu Zhang,
Xuanhua Shi,\\
Philip S. Yu~\IEEEmembership{Fellow,~IEEE,~ACM}

\IEEEcompsocitemizethanks{
    \IEEEcompsocthanksitem Yang Wu, Yao Wan and Zhaoyang Chu are with the Services Computing Technology and System Lab, Cluster and Grid Computing Lab, School of Computer Science and Technology, Huazhong University of Science and Technology, Wuhan, China, 430074. \protect 
    Email: \{wuyang\_emily,wanyao,chuzhaoyang,xhshi\}@hust.edu.cn. 
      \IEEEcompsocthanksitem 
    Wenting Zhao and Philip S. Yu are with the Big Data and Social Computing Lab, University of Illinois Chicago, Chicago, USA. \protect Email: \{wzhao41,psyu\}@uic.edu.
    \IEEEcompsocthanksitem 
    Ye Liu is with Salesforce Research, Palo Alto, USA. \protect Email: yeliu@salesforce.com.
    \IEEEcompsocthanksitem Hongyu Zhang is with the School of Big Data and Software Engineering, Chongqing University, Chongqing, China. \protect Email: hyzhang@cqu.edu.cn.
	}
    \thanks{*Yao Wan is the corresponding author.}
    \thanks{Manuscript received November, 2024.}
}

\maketitle

\begin{abstract}
Code summarization facilitates program comprehension and software maintenance by converting code snippets into natural-language descriptions. Over the years, numerous methods have been developed for this task, but a key challenge remains: effectively evaluating the quality of generated summaries.
While human evaluation is effective for assessing code summary quality, it is labor-intensive and difficult to scale. Commonly used automatic metrics, such as BLEU, ROUGE-L, METEOR, and BERTScore, often fail to align closely with human judgments.
In this paper, we explore the potential of Large Language Models (LLMs) for evaluating code summarization. We propose \system (Role-Player for Code Summarization Evaluation), a novel method that leverages role-player prompting to assess the quality of generated summaries. Specifically, we prompt an LLM agent to play diverse roles, such as code reviewer, code author, code editor, and system analyst. Each role evaluates the quality of code summaries across key dimensions, including coherence, consistency, fluency, and relevance.
We further explore the robustness of LLMs as evaluators by employing various prompting strategies, including chain-of-thought reasoning, in-context learning, and tailored rating form designs. The results demonstrate that LLMs serve as effective evaluators for code summarization methods. Notably, our LLM-based evaluator, \system, achieves an 81.59\% Spearman correlation with human evaluations, outperforming the existing BERTScore metric by 17.27\%.
\end{abstract}

\begin{IEEEkeywords}
Code Summarization, Large Language Models, Role Player, Model Evaluation
\end{IEEEkeywords}

\section{Introduction}
Code summarization, which summarizes code snippets into natural-language descriptions, plays an important role in program comprehension and software maintenance~\cite{iyer2016summarizing,leclair2019neural}.
Current approaches to code summarization heavily leverage techniques from Natural Language Processing (NLP), with the aim of translating code snippets from one linguistic representation to another.
Recently, this trend has been further bolstered by the emergence of Large Language Models (LLMs)~\cite{geng2024large}, e.g., GPT-4~\cite{openai2023gpt4blog}, Gemini~\cite{team2023gemini}, and Llama~\cite{touvron2023llama}, owing to their remarkable capabilities in NLP.

Despite significant advancements in generating code summaries, methods for evaluating their quality have not kept pace. 
Human evaluation remains the gold standard for assessing code summaries; however, it is labor-intensive and difficult to scale.
Previously, in alignment with the NLP research trajectory, numerous automatic metrics, such as BLEU~\cite{papineni2002bleu}, ROUGE~\cite{lin2004rouge}, METEOR~\cite{banerjee2005meteor}, and BERTScore~\cite{zhang2019bertscore}, have been widely adopted for evaluating code summarization models. 
These metrics assess model performance by automatically measuring the similarity between generated summaries and reference (or gold-standard) summaries, which are typically derived from comments provided by developers.
However, studies have shown that these automated metrics exhibit a relatively low correlation with human evaluations~\cite{shi2022evaluation,roy2021reassessing}.

Inspired by the remarkable capabilities of LLMs in understanding and generating natural language, recent studies propose leveraging LLMs directly as \textit{reference-free} evaluators~\cite{wang2023chatgpt,liu2303g,chiang2023can,chiang2023closer,chan2023chateval,kocmi2023large,zeng2023evaluating,zhuo2024ice}. This research rests on the premise that LLMs can assess candidate outputs based on their generation probabilities without requiring a reference summary, positing that LLMs can assign higher probabilities to outputs that are both fluent and of high quality. 
Building on these insights, this paper aims to investigate the following question: \textit{Can LLMs effectively serve as evaluators for code summarization?}

\begin{figure}[!t]
\centering
\includegraphics[width=.45\textwidth]
{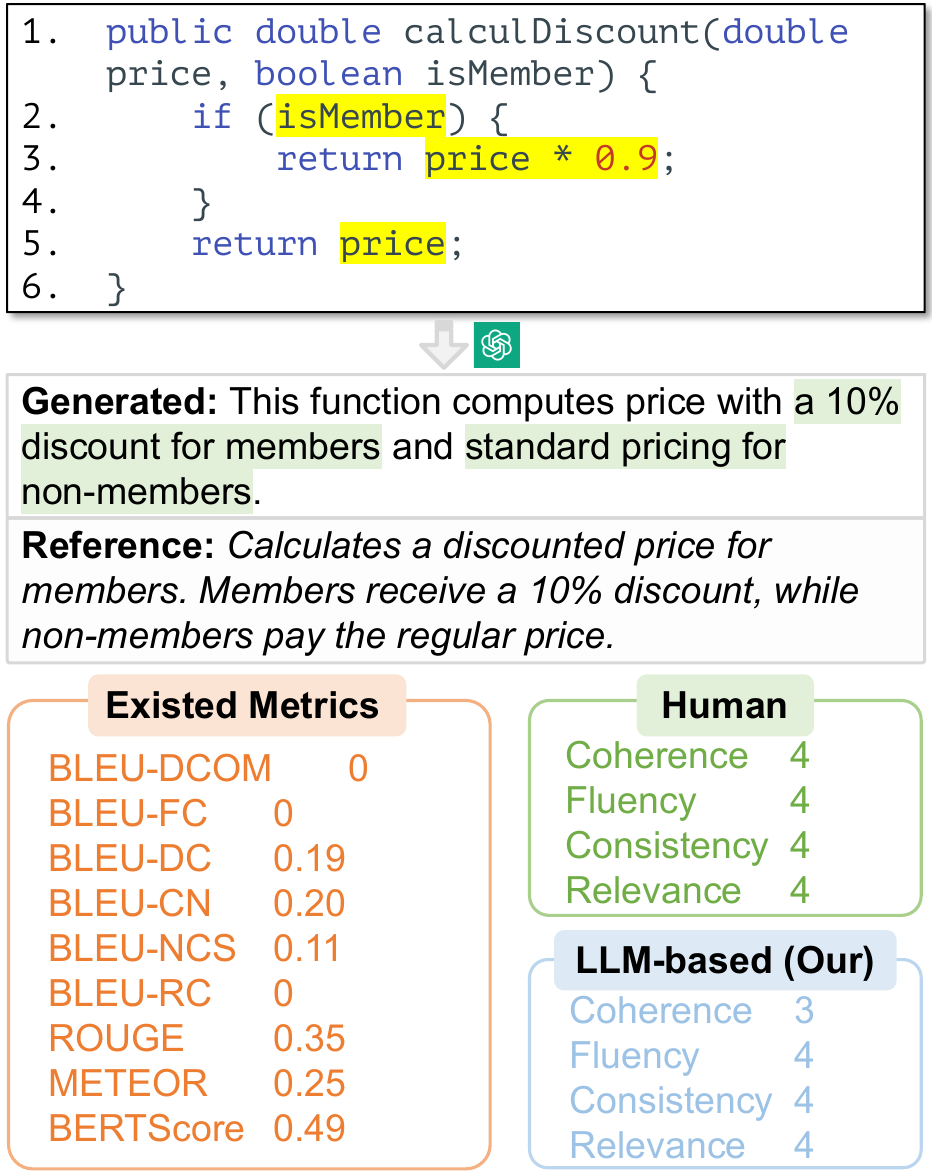}
\caption{A code snippet with reference and generated summaries, along with scores from existing metrics, human evaluation, and LLMs.}
\label{fig_example}
\end{figure}

\mypara{A Motivating Example}
Figure~\ref{fig_example} illustrates an example aimed at providing motivation for our introduced LLM-based evaluator by comparing it with current automatic metrics. 
In this example, a code snippet accompanied by its corresponding comment, alongside the summary generated by ChatGPT~\cite{openai2022chatgpt}, is presented. Upon employing automatic metrics such as BLEU, ROUGE, METEOR, and BERTScore, it is evident that the scores are relatively low, suggesting subpar quality in the generated summary. Nevertheless, upon manual inspection, the quality of the summary generated by ChatGPT is 
actually high.
Specifically, the phrase ``\textit{a 10\% discount for members}'' in the generated summary correlates directly to the variable ``\texttt{isMember}'' at line 2 and ``\texttt{0.9}'' at line 3 in the provided code. 
This can also reveal the strong understanding and reasoning capabilities of LLMs over source code.
We attribute the low scores from automatic metrics to the differences in textual structure and phrasing between the generated summary and the reference summary.
Conversely, our LLM-based evaluator showcases its ability to effectively assess the quality of generated summarizations, demonstrating significant alignment with human evaluation metrics.

\mypara{Our Work}
In this paper, we present a pioneering empirical study aiming at investigating the capabilities of LLMs in evaluating code summarization models. 
In particular, we introduce an evaluation method, termed \system (Role-Player for Code Summarization Evaluation), designed to quantify the quality of generated code summaries. 
We prompt an LLM agent to perform a range of roles, including code reviewer, code author, code editor, and system analyst. Each role is tasked with evaluating the quality of generated code summaries along individual dimensions such as coherence, consistency, fluency, and relevance. By analyzing the outcomes, we determine the most proficient role for each dimension to ensure a more precise and specialized evaluation.
In our experiments, we concentrate on three specific LLMs: \texttt{text-davinci-003}~\cite{openai2023gpt35}, \texttt{gpt-3.5-turbo}~\cite{openai2023gpt35}, and \texttt{gpt-4}~\cite{openai2023gpt4}. We assess the performance of six code summarization models, namely CodeNN~\cite{iyer2016summarizing}, Deepcom~\cite{hu2018deep}, Astattgru~\cite{leclair2019neural}, Rencos~\cite{zhang2020retrieval}, NCS~\cite{ahmad2020transformer}, and ChatGPT~\cite{sun2023automatic}.
We structure our empirical study around the following three Research Questions (RQs).

\smallskip
\noindent\textbf{RQ1: 
How does LLM-based evaluator \system align with human evaluation compared to traditional metrics?
}
We explore LLMs' capabilities in assessing code summarization, both with and without reference summaries, by examining their alignment with human evaluation alongside conventional metrics such as BLEU, ROUGE, METEOR, and BERTScore.

\noindent\textbf{RQ2: 
How does the LLM-based evaluator \system perform under varying evaluation settings?
}
We analyze different prompting strategies that may affect the performance of LLMs in evaluating code summarization, including the role-player design, rating form design, chain-of-thought prompting, and in-context example selection.
Our analysis provides clear guidelines for employing LLMs to automate the evaluation of code summarization.

\noindent\textbf{RQ3: 
How do existing neural models for code summarization perform using our proposed \systemnospace?
}
We re-evaluate the effectiveness of six prominent neural models for code summarization tasks, namely CodeNN, Deepcom, Astattgru, Rencos, NCS, and ChatGPT, utilizing our LLM-based evaluator, \systemnospace.

\mypara{Takeaway Implications}
In this paper, we outline several important implications to be noted:
(1) Overall, our  
\system demonstrates notable effectiveness in serving as evaluators for code summarization, exhibiting a correlation of 81.59\% 
to human assessment, even in the absence of reference summaries.
(2) The effectiveness of LLMs in assessing code summarization relies heavily on carefully crafted role-player engagement and prompting strategies. We recommend integrating a balanced selection of in-context learning examples and increasing evaluation iteration frequency. Additionally, employing chain-of-thought processes can significantly enhance fluency assessment.
(3) With our LLM-based evaluator \systemnospace, ChatGPT outperforms other models by producing summaries that maintain semantic accuracy while offering diverse phrasing, closely aligning with human evaluations.

\mypara{Contributions}
This paper makes the following contributions.
\begin{itemize}[leftmargin=4mm, itemsep=0.05mm]
    \item 
    To the best of our knowledge, we perform a pioneering investigation into the ability of LLMs to assess code summarization. 
    Furthermore, we introduce a novel evaluation approach called \system which leverages a roleplayer-based prompting strategy to evaluate the coherence and effectiveness of generated summaries, based on an understanding of the code itself, rather than relying on reference ground truth.
    \item 
    We extensively conduct experiments to compare \system with existing metrics in evaluating neural code summarization, employing various prompting strategies. Our experimental findings demonstrate that the LLM-based evaluator substantially enhances the correlation with human judgement across multiple criteria.
    \item  
    We reassess the efficacy of six prominent neural models (i.e., CodeNN, Deepcom, Astattgru, Rencos, NCS, and ChatGPT) in the realm of code summarization, leveraging our novel LLM-based evaluation framework, \systemnospace.
\end{itemize}

\section{Background}
\begin{figure*}[!t]
\centering
\includegraphics[width=0.98\textwidth]
{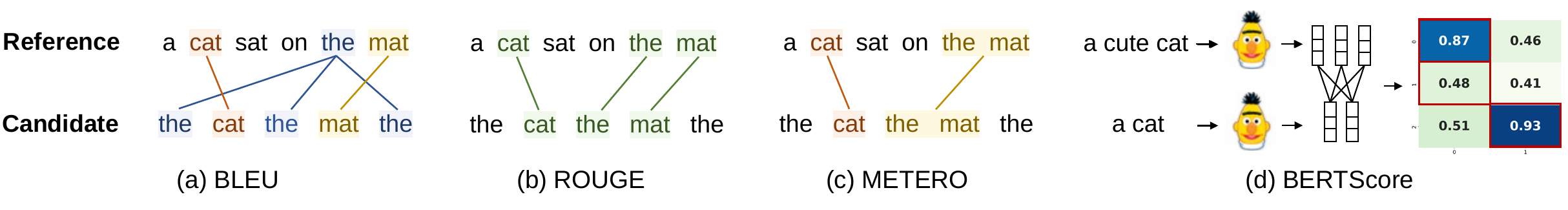}
\caption{An illustration of existing evaluation metrics for code summarization.
}
\label{fig_metrics}
\end{figure*}

\subsection{Existing Code Summarization Evaluation Metrics}
We classify prevailing evaluation metrics for code summarization into three categories: $n$-gram-based metrics, embedding-based metrics, and human evaluation metrics.

\subsubsection{$N$-gram-based Metrics} 
The objective of $n$-gram-based metrics is to calculate the similarity between the generated summary and the reference summary through the counting of shared $n$-grams. 
Examples include BLEU~\cite{papineni2002bleu}, ROUGE-L~\cite{lin2004rouge}, and METEOR~\cite{banerjee2005meteor}.

\mypara{BLEU~\cite{papineni2002bleu}}
BLEU (Bilingual Evaluation Understudy) is a traditional neural machine translation metric that calculates average $n$-gram precision with reference sentences and penalizes overly short translations.
\begin{equation}
    p_n = \frac{\sum_{w_n\in c} \min \left ( C_c(w_n), \underset{j=1,\cdots,n}{\max}C_{r_j}(w_n) \right )}{\sum_{w_n\in c} C_c(w_n)}\,,
  \end{equation}
where \( c \) and \( r \) represent a candidate and its reference sentence, respectively. \( w_n \) is an \( n \)-gram, and \( C_c(w_n) \) denotes its frequency in \( c \). The BLEU score is then computed as:
\begin{equation}
    \operatorname{BLEU} = \operatorname{BP}* \exp \left(\sum_{n=1}^{N}\alpha_n\log p_n\right)\,,
\end{equation}
where $N=4$, $p_n$ denotes the precision for $n$-grams up to $N$, and $\alpha_n$ represents the positive weighting assigned to each $n$-gram, and a brevity penalty BP penalizes the short sentences.

In practice, various implementations of the BLEU exist, each with specific modifications to handle $n$-gram precision differently. 
BLEU-CN is a sentence BLEU metric applying the Laplace-like smoothing~\cite{chen1999empirical} to the precision scores $p_n$ for $n \geqslant 2$, by both the numerator and denominator by 1. 
BLEU-DM and BLEU-DC are two metrics that apply another smoothing method which has been implemented in the NLTK toolkit\footnote{https://pypi.org/project/nltk/3.2.4/} as ``method 0'', and ``method 4'', respectively.
Similar to BLEU-CN, BLEU-NCS~\cite{ahmad2020transformer} applies a Laplace-like smoothing method, incrementing the numerator and denominator of all precision values $p_n$ by 1.
BLEU-RC~\cite{zhang2020retrieval} is another unsmoothed sentence BLEU variant designed specifically to prevent division-by-zero errors. Instead of traditional smoothing, it adds between 10 and 15 to the numerator, and between 9 and 10 to the denominator of $p_n$.
BLEU-FC is an unsmoothed corpus-level BLEU metric based on NLTK, which aggregates n-gram matches across all hypothesis-reference pairs~\cite{leclair2019neural,wei2020retrieve}. 

\noindent\underline{\textsc{\textbf{Example 1.}}} 
In the example illustrated in Figure~\ref{fig_metrics}(a) for calculating BLEU, the candidate sentence yields unigram counts of ``the", ``cat", and ``mat" as 3, 1, and 1, respectively, with a total count of 5. 
For reference, the relevant unigrams ``the," ``cat," and ``mat" each have a count of 1. By taking the minimum counts between the candidate and reference, we sum to 3. 
Thus, with $n=1$, we have $p_1=0.6$, yielding a BLEU score of 0.49. 

\mypara{ROUGE-L~\cite{lin2004rouge}}
ROUGE-L~\cite{bansal2021project,lin2021improving,shahbazi2021api2com} quantifies the similarity between a candidate summary and a reference summary by identifying their Longest Common Subsequence (LCS). This metric emphasizes the importance of maintaining the sequential order of information. The calculation of the ROUGE-L score and its components is defined as follows:
\begin{equation} 
    \begin{split}
    & P = \frac{LCS(c, r)}{len(c)}, R = \frac{LCS(c, r)}{len(r)} \\
    & \operatorname{ROUGE-L} = \frac{(1 + \beta^2) \cdot P \cdot R}{R + \beta^2 \cdot P}\,.
    \end{split}
\end{equation}
Here, \(c\) and \(r\) denote the generated candidate and reference summaries, respectively, and \(\beta\) is a parameter that adjusts the balance between precision and recall. 

\noindent\underline{\textsc{\textbf{Example 2.}}} As the example shown in Figure~\ref{fig_metrics} (b), with consideration of the longest common sequence of words is ``cat the mat'', we can get $LCS(c,r)=3,\ \text{then} \ P=0.6,\ \text{and}\ R=0.5$. When $\beta=1$, $\text{ROUGE-L}$ equals to 0.54.

\mypara{METEOR}
METEOR, a recall-oriented metric, evaluates how well the model captures reference content by matching words between candidate and reference sentences and computing the harmonic mean of precision and recall, calculated as follows:
\begin{multline}
    \operatorname{METEOR} = \underset{j=1,\cdots,n}{\max}\left ( \frac{10PR}{R+9P} \right )\left ( 1-\frac{1}{2}  (\frac{c}{u})^3 \right )\,,
\end{multline}
where \( P \) and \( R \) denote unigram precision and recall, \( c \) is the count of matched chunks, and \( u \) represents matched unigrams.

\noindent\underline{\textsc{\textbf{Example 3.}}} 
As illustrated in Figure~\ref{fig_metrics} (c), the alignment between the candidate and reference sentences produces two coherent chunks: ``cat'' and ``the mat'', resulting in a total of three matched unigrams. With a precision ($P$) of 0.6 and recall ($R$) of 0.5, the METEOR score is calculated as 0.5.

\subsubsection{Embedding-based Metrics}
While $n$-gram-based metrics assess semantic similarity through overlapping tokens, embedding-based metrics like BERTScore~\cite{zhang2019bertscore} compare the embeddings of generated and reference summaries.

\mypara{BERTScore~\cite{zhang2019bertscore}}
BERTScore measures sentence similarity using BERT embeddings~\cite{zhang2019bertscore}, matching each token in the candidate sentence to tokens in the reference. It tokenizes sentences into words or subwords and represents tokens as embeddings using a pre-trained BERT model, e.g., RoBERTa~\cite{liu2019roberta}.
Specifically, BERTScore computes a pairwise similarity matrix by taking the inner product of these embeddings, yielding a pre-normalized cosine similarity. Precision is defined as the average of the highest similarity scores for tokens in the candidate sentence, while recall is the average of the highest scores for tokens in the reference sentence. The final BERTScore is computed as the harmonic mean of precision and recall.

\noindent\underline{\textsc{\textbf{Example 4.}}} 
In Figure~\ref{fig_metrics}(d), using the RoBERTa model, we compute a cosine similarity matrix between the words of the candidate and reference sentences. 
Precision is computed by averaging the maximum similarity scores for each word in the candidate, yielding a value of $(0.87 + 0.93) / 2 = 0.90$. Recall is calculated by averaging the maximum similarity scores for each word in the reference, resulting in $(0.87 + 0.48 + 0.93) / 3 = 0.76$.
The $\text{BERTScore}$, computed as the harmonic mean of a precision of 0.9 and a recall of 0.76, yields a value of 0.82.

\begin{figure}[!t]
\centering
\includegraphics[width=0.5\textwidth]
{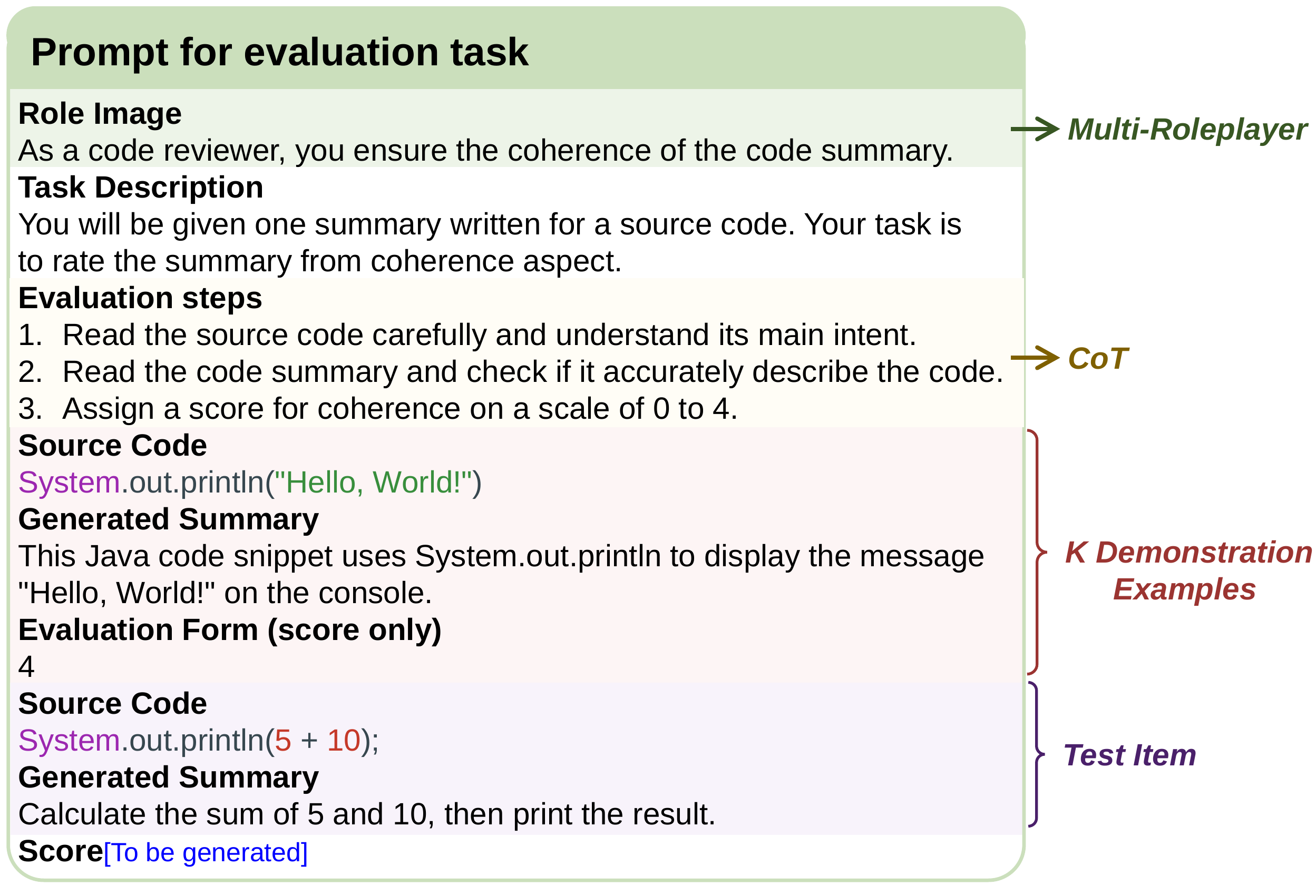}
\caption{An example of the prompt design to elicit LLMs as a code evaluator taking a ``\textit{code reviewer}'' role.
}
\label{fig_llms}
\end{figure}

\begin{figure*}[!t]
\centering
\includegraphics[width=.98\textwidth]
{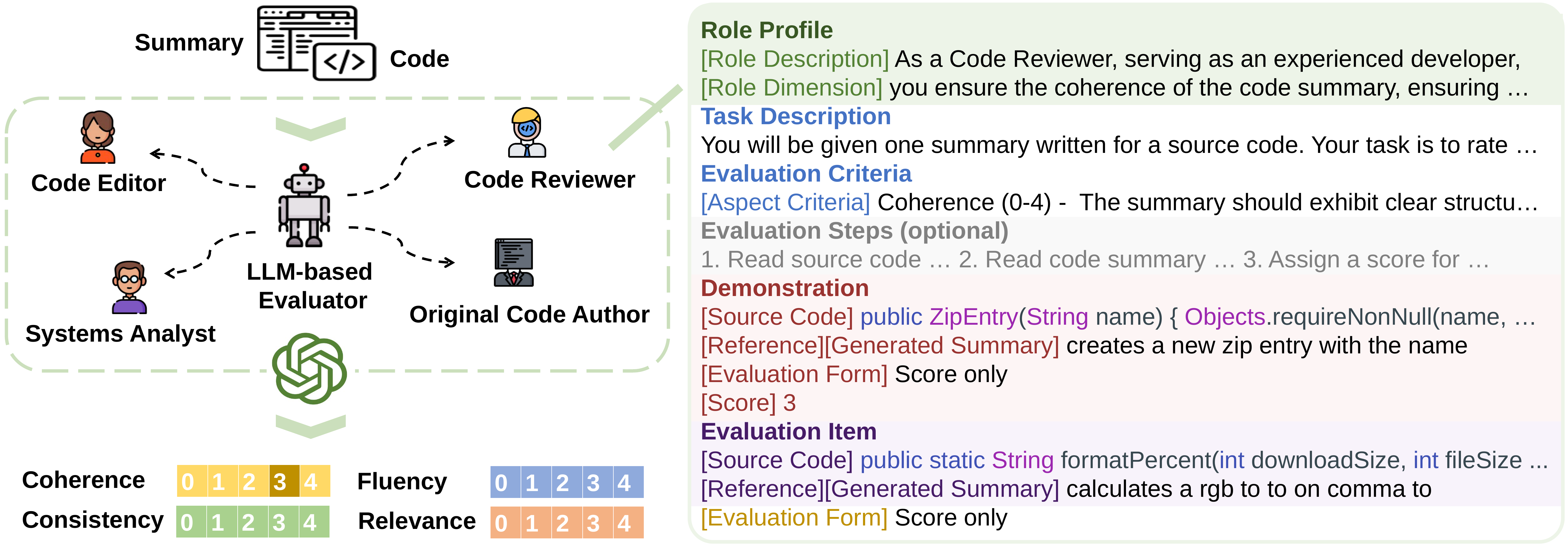}
\caption{An overview of our proposed framework for prompting LLMs as diverse evaluators for code summarization. 
}
\label{fig_overview}
\end{figure*}

\subsubsection{Human Evaluation Metrics}
In human evaluation, human assessors are typically presented with pairs of references and model-generated outputs. The assessors then evaluate the outputs based on predefined criteria such as fluency, coherence, and overall understanding. Often, a 5-point scale, ranging from 1 to 5 with options like ``Strongly Disagree'', ``Disagree'', ``Neutral'', ``Agree'', and ``Strongly Agree'' is employed, allowing assessors to express their judgments on a numerical scale. 
Aggregating these human judgments yields scores that reflect the perceived quality of the generated text. 
However, the expense associated with hiring professional evaluators makes human evaluation difficult to scale. Furthermore, human evaluation metrics introduce subjectivity, necessitating efforts to mitigate bias and ensure consistency in the evaluation process.

\subsection{Large Language Models}

\mypara{Prompting}
The advent of LLMs is shifting the learning paradigm from the traditional ``pre-train and fine-tune'' to a novel ``pre-train, prompt, and predict'' approach. In this framework, downstream tasks are reformulated using textual prompts to align with the original LLM training, rather than through full fine-tuning.
In our specific scenario, involving a summary and a code snippet, we can construct a prompt as follows: ``\texttt{[SUMMARY]}, \texttt{[SOURCE CODE]}, \textit{Please rate the coherence of the generated summary based on the source code}'', where \texttt{[SUMMARY]} signifies the model-generated summary, and \texttt{[SOURCE CODE]} represents the code snippet.

\mypara{In-Context Learning (ICL)}
ICL is a special form of prompt-based learning that leverages demonstration examples in prompts to promote the model's performance.
Specifically, given a test question $x_t$, ICL retrieves $k$ examples related to $x_t$ from the task dataset as demonstrations and uses the prompt function $f$ to transform these examples, creating a set of demonstration examples $D_{k}=\{f(x_{i},y_{i}),\ldots,f(x_{k},y_{k})\}$. Then, the LLMs predict $\hat{y}_{t}$ based on the task description $I$ and example set $D_k$.

\mypara{Chain-of-Thought (CoT)}
CoT~\cite{wei2022chain} is a prompting technique that enhances LLM performance on complex reasoning tasks by incorporating intermediate reasoning steps into ICL prompts to guide the model toward the final output.
Specifically, the CoT prompting strategy augments each demonstration example $\langle x, y\rangle$ in ICL with a chain-of-thought prompt $CoT$, constructing a triplet prompt $\langle x, CoT, y \rangle$.

\mypara{Multi-Role Player}
LLMs have demonstrated the capability to simulate human behavior through role-playing, adapting to diverse roles across contexts. Recent studies leverage this versatility, from simulating character personalities in gaming~\cite{park2023generative} to aiding consensus-building in robotics~\cite{chen2023multi}, and facilitating debate evaluations in multi-agent systems~\cite{chan2023chateval}. In this work, we engage LLMs in expert roles to elicit their ability in code summary evaluation.

\noindent\underline{\textsc{\textbf{Example 5.}}}
Figure~\ref{fig_llms} showcases a prompt design that casts an LLM in the role of a ``\textit{code reviewer}''. The figure details five key components of LLMs: 1) a role, which assigns the expected identity and function of the LLM during the task, 2) a task description that clearly articulates the goal, 3) evaluation steps that segment the task into manageable parts, 4) $k$ demonstration examples that provide a model for performing the task, and 5) a test example for the LLM to assess.

\section{LLMs as Code Summarization Evaluators} 
This section presents \systemnospace, a novel evaluation method that employs a role-player prompting strategy to assess the quality of generated summaries, accommodating both reference-based and reference-free scenarios, as shown in Figure~\ref{fig_overview}.

\subsection{Multi-Role Player Design}
\label{multirole}
The core idea of \system is designing a multi-role player prompting strategy.
Specifically, we prompt an LLM agent to play diverse roles such as code reviewer, code author, code editor, and system analyst. 
Each role is required to evaluate the quality of generated code summaries, focusing on one dimension at a time,
such as coherence, consistency, fluency, and relevance. 
The schema of \textit{role profile prompt} is presented as \textit{$\langle role\ description,\ role\ dimension \rangle$}. 
We replace the \textit{role description} slot with a variety of roles (e.g., code reviewer, original code author, code editor, and system analyst) and investigate their performance on the specific role dimension. 
The slot of \textit{role dimension} could be filled with descriptions of various dimensions (e.g., coherence, consistency, fluency, and relevance) to guide the roles to behave. The specific descriptions are as follows. 
\begin{tcolorbox}
[left=1pt,right=1pt, top=1pt, bottom=1pt,colback=white, colframe=black,boxrule=0.8pt]
\noindent\textit{\textbf{Coherence Dimension}}: Ensure that the summary captures the primary functionality and logic of the code without introducing any additional or unrelated content.
\end{tcolorbox}
\begin{tcolorbox}
[left=1pt,right=1pt, top=1pt, bottom=1pt,colback=white, colframe=black,boxrule=0.8pt]
\noindent\textit{\textbf{Consistency Dimension}}: Guarantee that the summary remains consistent with the original code, accurately capturing its primary functionality and logic without adding any unrelated content.

\noindent\textit{\textbf{Fluency Dimension}}: Focus on ensuring that the summary is written smoothly, with clear sentences and appropriate wording.

\noindent\textit{\textbf{Relevance Dimension}}: Concentrate on the business or functional relevance of the code, ensuring that the summary captures the key significance of the code within the larger system or project.
\end{tcolorbox}

\subsection{Prompting Strategies}
\label{sec_prompt_stragety}
To clarify the evaluation task for LLMs, we provide a \textit{task description} stating, ``\textit{you will be given one summary written for a source code and your task is to rate the summary on one metric}''. 
Subsequently, we investigate various prompt strategies to enhance the performance of LLMs.
\label{strategies}

\begin{tcolorbox}
[left=1pt,right=1pt, top=1pt, bottom=1pt,colback=white, colframe=black,,boxrule=0.8pt]
1. Read the source code carefully and understand its main functionality and key operations.

2. Read the code comments and compare them to the source code. Check if the comments accurately describe the main functionality and key operations of the code, and present them in a clear and logical order. 

3. Assign a score for coherence on a scale of 0 to 4, based on the Evaluation Criteria. 
\end{tcolorbox}

\subsubsection{CoT}
\label{cot}

Considering the evaluation task, 
detailed evaluation instructions can guide the annotator in inferring the rating score.
We follow the CoT strategies~\cite{liu2303g} to prompt LLM to generate detailed evaluation steps on its own.
Specifically, we incorporate the \textit{evaluation step} illustrated in Figure~\ref{fig_overview}, detailing the reasoning process to derive the final score. The specific stages within the evaluation step are outlined above.

\subsubsection{ICL} 
\label{icl}
To enhance LLMs' performance in the evaluation task, we provide them with selected annotated examples. As illustrated in Figure~\ref{fig_overview}, we employ a \textit{demonstration prompt}, which can be extended by adjusting the number of examples, denoted as $K$. Each example is meticulously linked to a specific evaluation dimension and rated on a scale of 0 to 4. These examples comprise a code snippet,  an optional reference summary, a generated summary, a specific rating form, and an associated human-assigned score. 

\subsubsection{Rating Forms}
\label{rating_forms}
Lastly, we feed LLMs the generated summary for evaluation, along with its source code and an optional reference summary, ending with a \textit{evaluation form} slot in the \textit{evaluation item prompt}. We explore diverse scoring formats to guide LLMs to output ratings. In addition to the traditional \textit{score-only form}, where LLMs output only a numerical score, we introduce two more nuanced approaches based on the evaluation guidelines detailed in~\cite{chiang2023closer}: the \textit{analyze-rate form} and the \textit{rate-explain form}. The \textit{analyze-rate form} requires LLMs to process the reasoning behind their assessment before giving a score, whereas the \textit{rate-explain form} prompts LLMs to score first and then justify their evaluation. Specifically, we replace the \textit{evaluation form} slot with diverse form descriptions as follows.

\begin{tcolorbox}
[left=1pt,right=1pt, top=1pt, bottom=1pt,colback=white, colframe=black,,boxrule=0.8pt]
\noindent\textit{\textbf{Score-only Form}}: 
``\textit{Score only}''.

\noindent\textit{\textbf{Analyze-rate Form}}: 
``\textit{Answer by starting with `Analysis' to analyze the given example regarding the evaluation criteria as concisely as possible, and then give the numeric rating on the next line by `Rating'}''.

\noindent\textit{\textbf{Rate-explain Form}}: 
``\textit{Answer by starting with `Rating' and then give the explanation of the rating on the next line by `Rationale'}''.
\end{tcolorbox}

\begin{table*}[t!]
    \centering
    \caption{
The overall performance of code summarization by employing
\system backend by \texttt{gpt-4} across various role players, compared with
conventional metrics.
}
    \setlength{\tabcolsep}{8pt} 
    \begin{tabular}{l|cc|cc|cc|cc|cc}
        \hline
        \multicolumn{1}{l|}{\multirow{2}{*}{\centering \textbf{Metric}}}
        &\multicolumn{2}{c}{\textbf{Coherence}} & \multicolumn{2}{c}{\textbf{Consistency}} & \multicolumn{2}{c}{\textbf{Fluency}}&\multicolumn{2}{c}{\textbf{Relevance}} & \multicolumn{2}{c}{\textbf{Average}} \\
        &\multicolumn{1}{c}{\textbf{$\tau$}} & \multicolumn{1}{c|}{\textbf{$\rho$}} 
        &\multicolumn{1}{c}{\textbf{$\tau$}} & \multicolumn{1}{c|}{\textbf{$\rho$}} 
        &\multicolumn{1}{c}{\textbf{$\tau$}} & \multicolumn{1}{c|}{\textbf{$\rho$}} 
        &\multicolumn{1}{c}{\textbf{$\tau$}} & \multicolumn{1}{c|}{\textbf{$\rho$}} 
        &\multicolumn{1}{c}{\textbf{$\tau$}} & \multicolumn{1}{c}{\textbf{$\rho$}} \\
        \hline
        \multicolumn{7}{l}{\textbf{\textit{Existing Metrics}}}\\
        \hline
        BLEU-DM &25.58\%&	53.85\%&	30.3\%&	63.06\%&	24.41\%&	51.67\%&	30.40\%&	63.26\%&	\cellcolor{gray!25}27.67\%&	\cellcolor{gray!25}57.96\%\\
        BLEU-FC &25.56\%&	53.85\%&	30.30\%&	63.06\%&	24.41\%&	51.67\%&	30.40\%&	63.26\%&	\cellcolor{gray!25}27.67\%&	\cellcolor{gray!25}57.96\%\\
        BLEU-DC &\textbf{42.21\%}&	\textbf{59.50\%}&	\textbf{51.24\%}&	\textbf{70.40\%}&	\textbf{39.70\%}&	\textbf{56.50\%}&	\textbf{51.54\%}&	\textbf{70.63\%}&	\cellcolor{gray!25}\textbf{46.17\%}&	\cellcolor{gray!25}\textbf{64.26\%}\\
        BLEU-CN &38.08\%&	53.83\%&	45.58\%&	62.00\%&	36.76\%&	52.21\%&	45.40\%&	62.52\%&	\cellcolor{gray!25}41.46\%&	\cellcolor{gray!25}57.64\%\\
        BLEU-NCS &32.86\%&	46.93\%&	34.56\%&	48.83\%&	32.43\%&	46.68\%&	34.51\%&	48.78\%&	\cellcolor{gray!25}33.59\%&	\cellcolor{gray!25}47.81\%\\
        BLEU-RC &25.56\%&	53.85\%&	30.30\%&	63.06\%&	24.41\%&	51.67\%&	30.40\%&	63.26\%&	\cellcolor{gray!25}27.67\%&	\cellcolor{gray!25}57.96\%\\
        ROUGE-L	&33.80\%&	46.58\%&	47.39\%&	62.72\%&	30.65\%&	42.71\%&	47.86\%&	63.30\%&	\cellcolor{gray!25}39.93\%&	\cellcolor{gray!25}53.83\%\\ 
        METEOR	&38.10\%&	53.29\%&	52.11\%&	69.32\%&	34.44\%&	48.70\%&	53.16\%&	70.26\%&	\cellcolor{gray!25}44.45\%&	\cellcolor{gray!25}60.39\% \\ 
        BERTScore &\textbf{44.61\%}&	\textbf{59.52\%}&	\textbf{55.10\%}&	\textbf{70.59\%}&	\textbf{41.38\%}&	\textbf{56.03\%}&	\textbf{55.72\%}&	\textbf{71.14\%}&	\cellcolor{gray!25}\textbf{49.20\%}&	\cellcolor{gray!25}\textbf{64.32\%}\\ 
        \hline
        \multicolumn{7}{l}{\textbf{\textit{LLM-based Evaluators}}}\\
        \hline
        \multicolumn{7}{l}{\textbf{\textit{\underline{Reference-based}}}} \\ 
        
        Code Reviewer   &\textbf{58.51\%}&	\textbf{80.86\%}&	\textbf{60.29\%}&	\textbf{83.39\%}&	\textbf{52.19\%}&	80.29\%&	60.85\%&	84.38\%&	\cellcolor{gray!25}\textbf{57.96\%}&	\cellcolor{gray!25}\textbf{82.23\%}\\ 
        Original Code Author   &55.19\%&	77.92\%&	59.37\%&	82.71\%&	51.07\%&	80.79\%&	59.76\%&	83.23\%&	\cellcolor{gray!25}56.35\%&	\cellcolor{gray!25}81.16\%\\ 
        Code Editor   &57.07\%&	79.46\%&	59.26\%&	83.17\%&	52.02\%&	\textbf{81.8\%}&	60.54\%&	83.06\%&	\cellcolor{gray!25}57.22\%&	\cellcolor{gray!25}81.87\%\\ 
        Systems Analyst   &57.17\%&	79.71\%&	58.68\%&	82.01\%&	49.88\%&	77.98\%&	\textbf{61.69\%}&	\textbf{84.95\%}&	\cellcolor{gray!25}56.86\%&	\cellcolor{gray!25}81.16\%\\ 

        \multicolumn{7}{l}{\textbf{\textit{\underline{Reference-Free}}}}	\\ 
        Code Reviewer&59.34\%& 82.31\%&	\textbf{58.63\%}&	\textbf{82.68\%}&	50.98\%&	\textbf{80.48\%}&	\textbf{56.96\%}&	\textbf{80.89\%}&	\cellcolor{gray!25}\textbf{56.48\%}&	\cellcolor{gray!25}\textbf{81.59\%} \\
        Original Code Author&57.49\%& 81.03\%&	57.88\%&	82.16\%&	49.83\%&	78.56\%&	55.78\%&	80.07\%&	\cellcolor{gray!25}55.25\%&	\cellcolor{gray!25}80.46\%\\
        Code Editor&\textbf{60.77\%}&	\textbf{83.38\%}&	56.74\%&	80.38\%&	\textbf{51.34\%}&	80.40\%&	55.84\%&	79.22\%&	\cellcolor{gray!25}56.17\%&	\cellcolor{gray!25}80.85\%\\
        System Analyst&60.24\%&	83.09\%&	57.97\%&	81.54\%&	49.98\%&	78.54\%&	55.31\%&	78.61\%&	\cellcolor{gray!25}55.88\%&	\cellcolor{gray!25}80.45\% \\
        \hline
    \end{tabular}
    \label{tab_overview}
\end{table*}

\section{Experiments}
We assess the performance of LLMs as evaluators for the code summarization task by addressing the following \textit{Research Questions} (RQs):
\begin{itemize}[leftmargin=4mm, itemsep=0.05mm]
    \item \textbf{RQ1: Performance of \system Evaluator.} \textit{To what extent does the LLM-based evaluator \system align with human evaluation compared to traditional metrics?}
    \item \textbf{RQ2: Influence of Evaluator Settings.} 
    \textit{How does the LLM-based evaluator \system perform across different settings, such as evaluator types, number of demonstration examples, turns, and prompt strategies?}
    \item \textbf{RQ3: Re-Evaluation of Current Models.} 
    \textit{How do existing neural models for code summarization perform when evaluated using our proposed LLM-based \systemnospace?}
\end{itemize}

\subsection{Datasets and Models to Re-Evaluate}

\subsubsection{Datasets}
We conduct experiments using TL-CodeSum~\cite{hu2018summarizing}, a widely used dataset for code summarization, following~\cite{shi2022evaluation}. The dataset contains 87,136 Java code-summary pairs from 9,732 GitHub projects (2015–2016) with at least 20 stars and is split into training, validation, and test sets in an 8:1:1 ratio.
To evaluate the correlation between automated metrics and human judgments, we utilize the 300 annotated summaries provided by~\cite{shi2022evaluation} as the ground-truth labels.

\subsubsection{Models to Re-Evaluate}
\label{eval_models}
In our experiments, we re-evaluate the following six code summarization models:
\noindent\textbf{CodeNN}~\cite{iyer2016summarizing} is an early neural model for code summarization, which encodes source code into context vectors and then generates summaries using an attention mechanism.
\noindent\textbf{Deepcom}~\cite{hu2018deep} linearizes source code by traversing its abstract syntax tree and employs an LSTM-based model to translate the traversal into a summary. 
\noindent\textbf{Astattgru}~\cite{leclair2019neural} employs two recurrent neural networks to encode both the lexical and syntactic information of source code.
\noindent\textbf{NCS}~\cite{ahmad2020transformer} utilizes a Transformer-based model to generate summaries for code.
\noindent\textbf{Rencos}~\cite{zhang2020retrieval} incorporates similar code snippets retrieved from the training dataset to enhance the code summarization model.
\noindent\textbf{ChatGPT}  (e.g., \texttt{gpt-3.5-turbo})~\cite{sun2023automatic} denotes the ChatGPT-based model for code summarization via prompting. 
The re-evaluation results are shown in Sec.~\ref{result_RQ3}.

\begin{figure*}[!t]
    \begin{subfigure}[b][][c]{.19\textwidth}
	\centering
        \includegraphics[width=\linewidth]{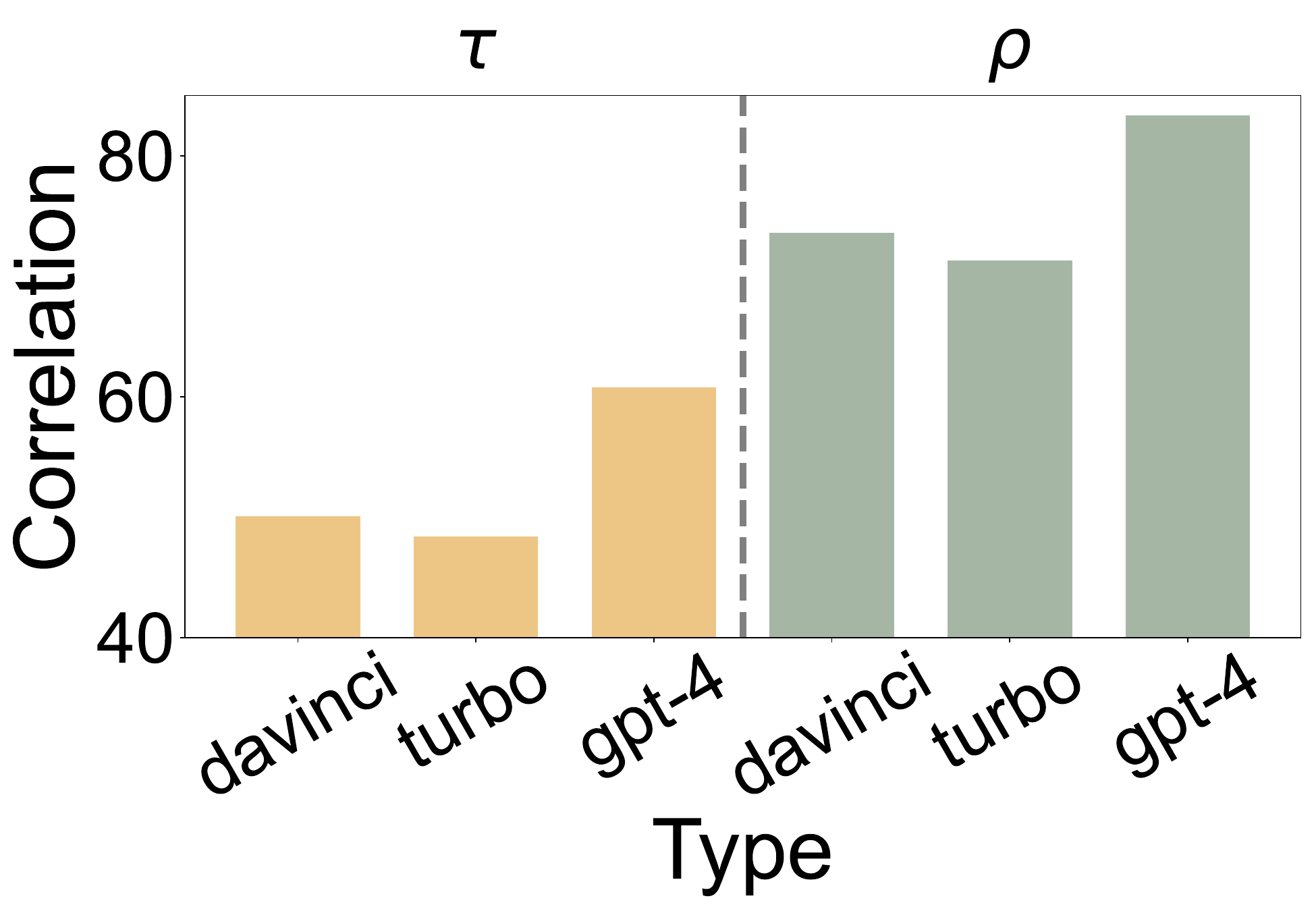}
	\caption{Coherence}
	\label{fig_type_coh}
    \end{subfigure}
    \begin{subfigure}[b][][c]{.19\textwidth}
	\centering
        \includegraphics[width=\linewidth]{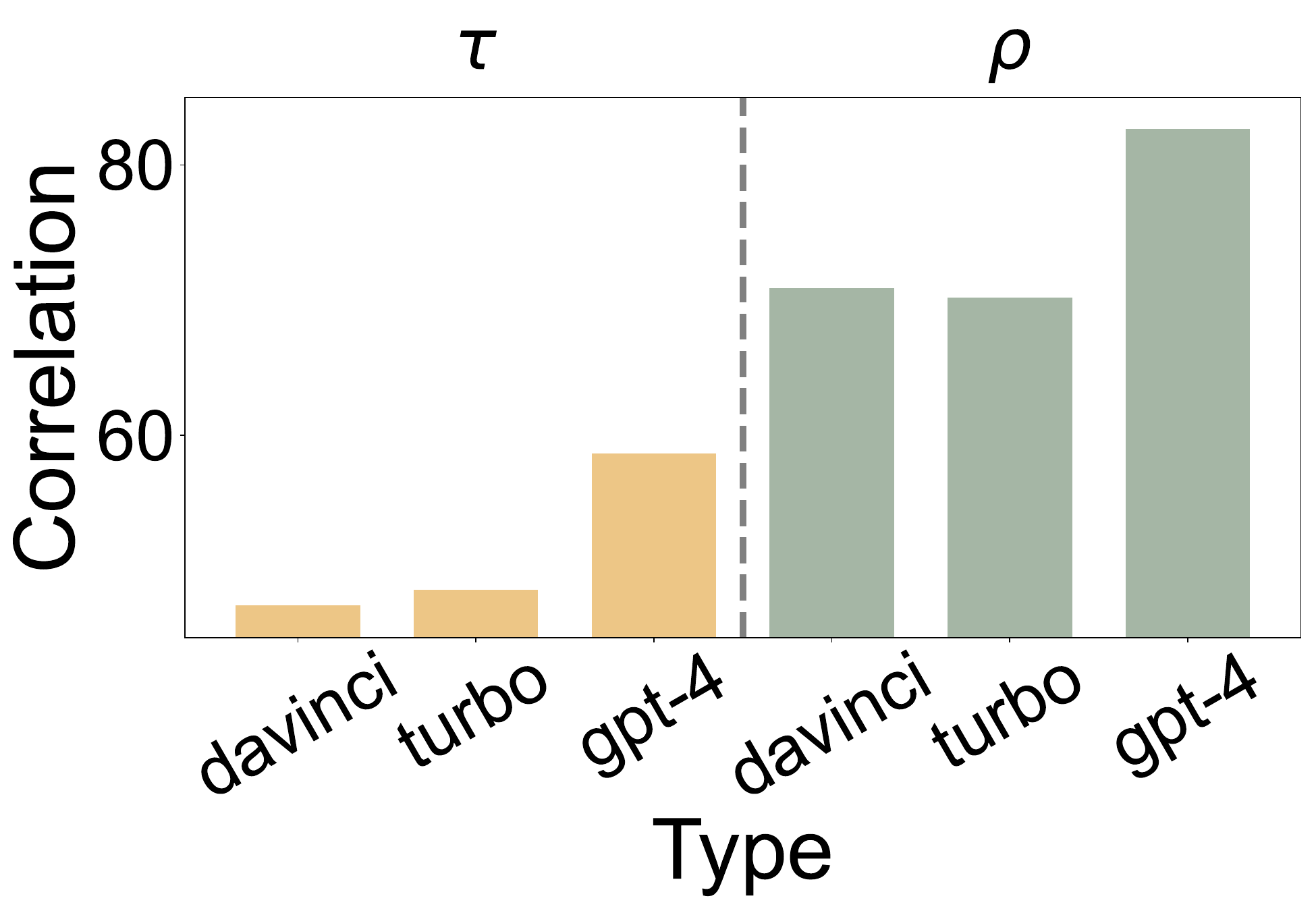}
	\caption{Consistency}
	\label{fig_type_con}
    \end{subfigure}
    \begin{subfigure}[b][][c]{.19\textwidth}
	\centering
        \includegraphics[width=\linewidth]{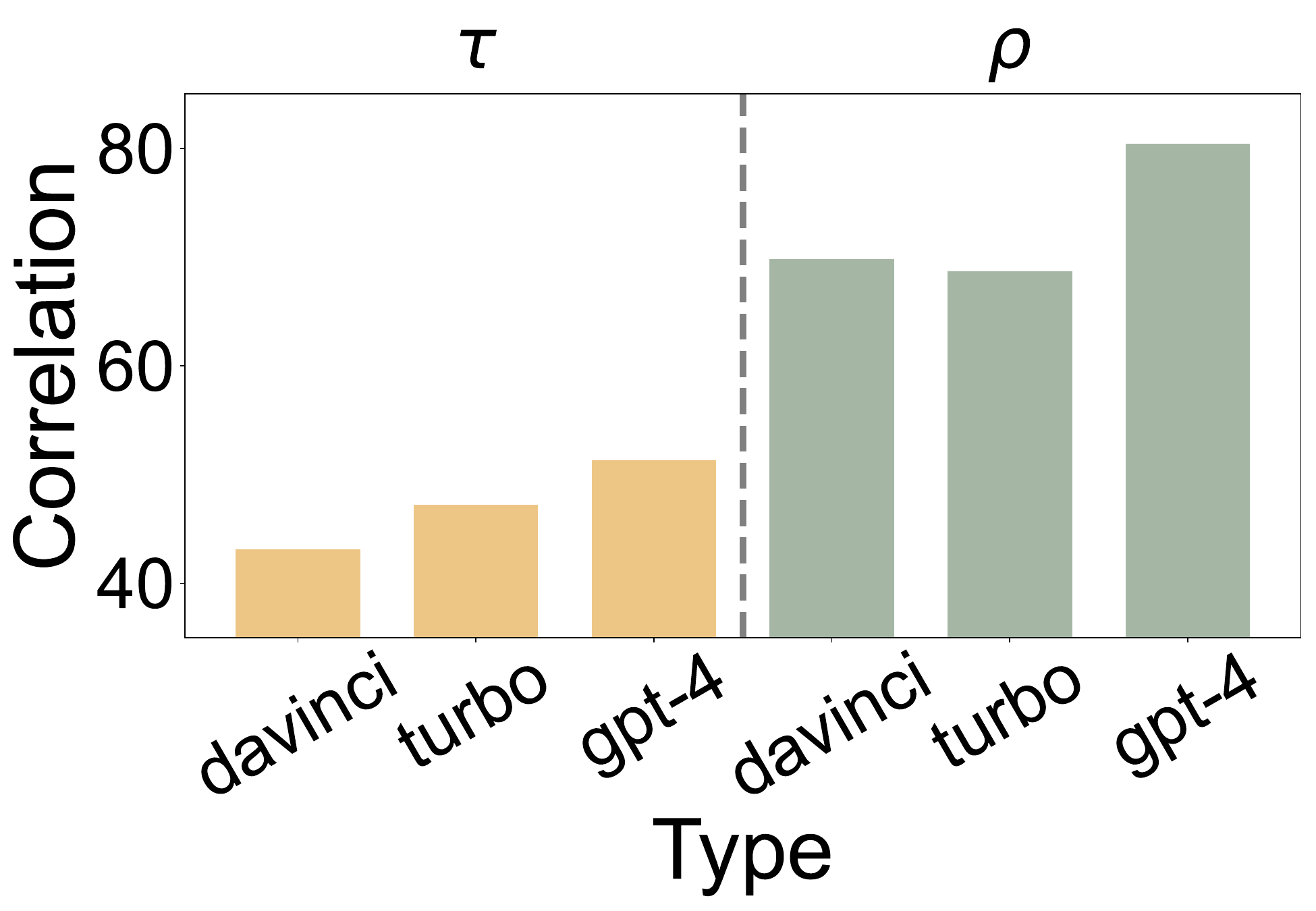}
	\caption{Fluency}
	\label{fig_type_flu}
    \end{subfigure}
    \begin{subfigure}[b][][c]{.19\textwidth}
	\centering
        \includegraphics[width=\linewidth]{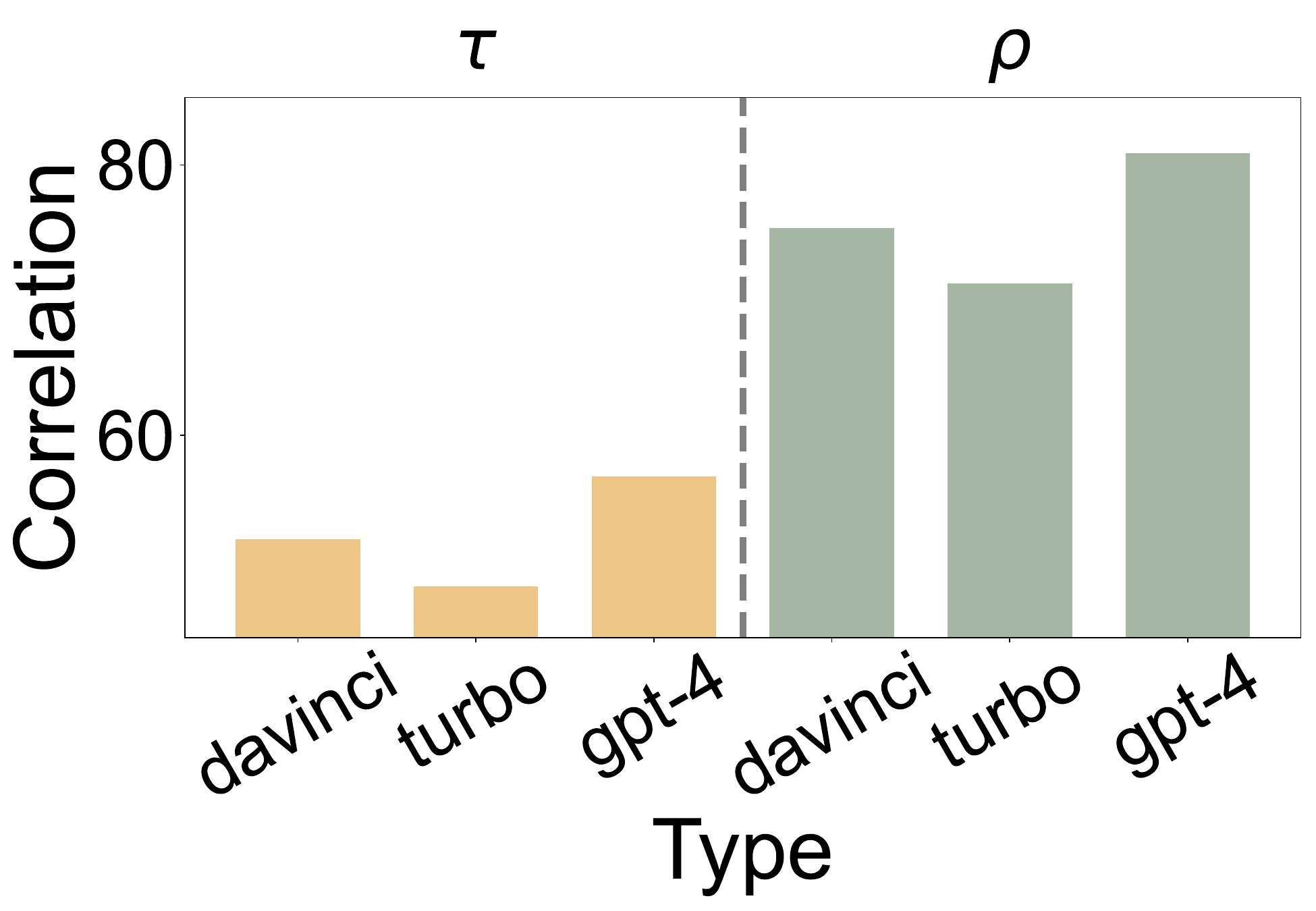}
	\caption{Relevance}
	\label{fig_type_rel}
    \end{subfigure}
    \begin{subfigure}[b][][c]{.19\textwidth}
	\centering
        \includegraphics[width=\linewidth]{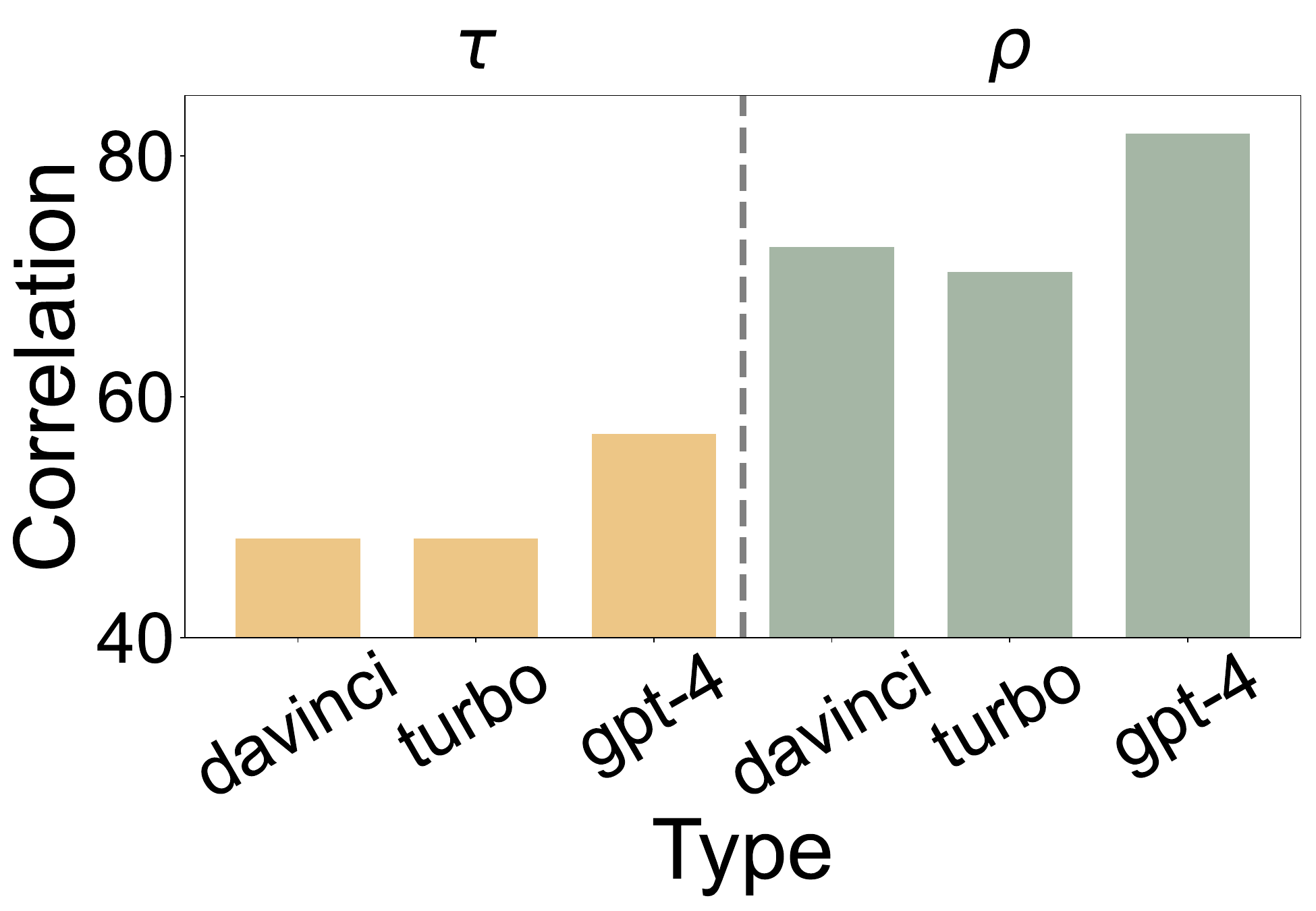}
	\caption{Average}
	\label{fig_type_avg}
    \end{subfigure}
\caption{Performance of evaluators across various LLMs in the reference-free scenario.
Here, ``davinci'' stands for ``\texttt{text-davinci-003}'', ``turbo'' represents ``\texttt{gpt-3.5-turbo}'', and ``\texttt{gpt-4}'' remains unchanged.
}
\label{figure_type}
\vspace{-4mm}
\end{figure*}

\begin{figure*}[!t]
    \begin{subfigure}[b][][c]{.19\textwidth}
	\centering
        \includegraphics[width=\linewidth]{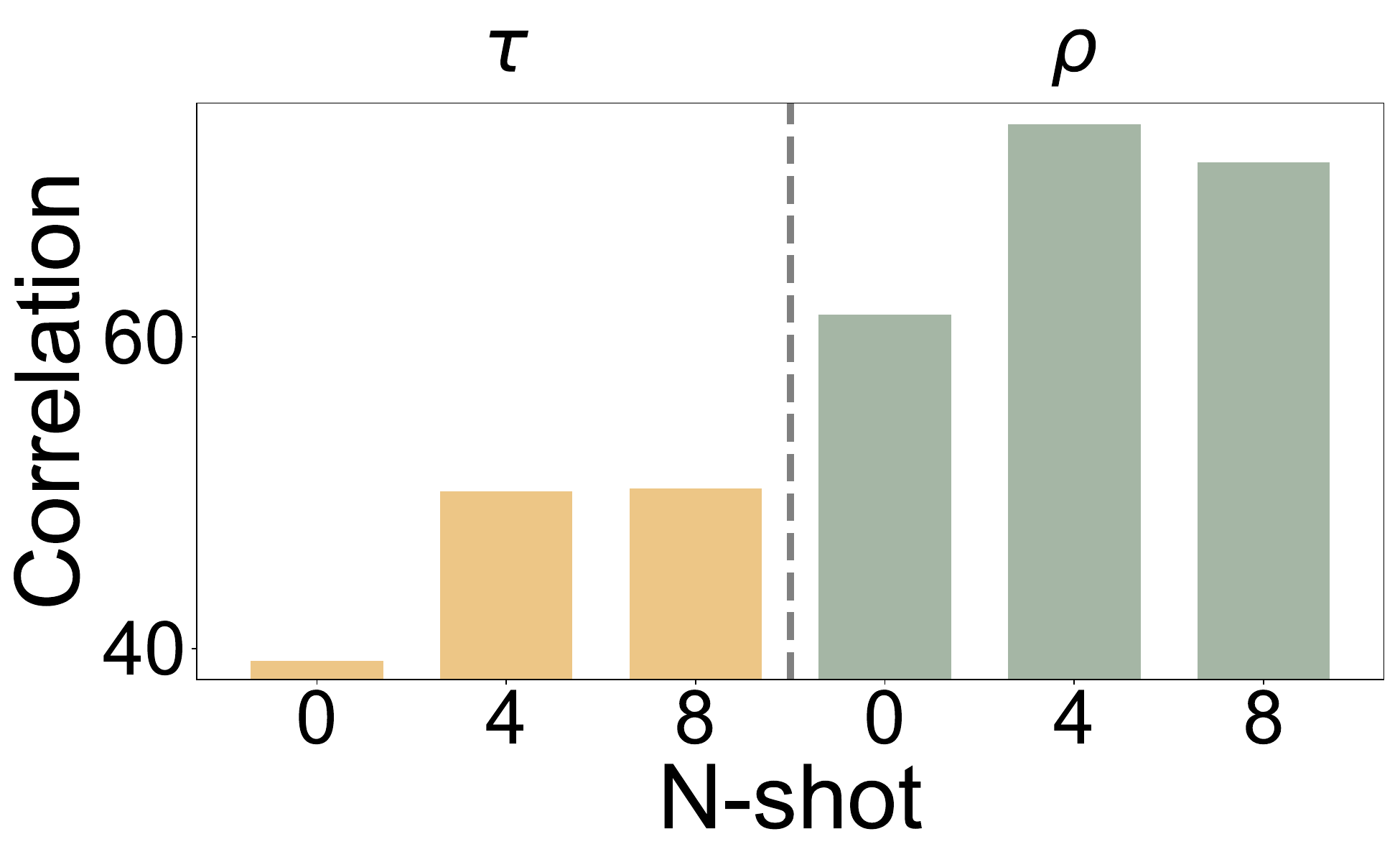}
	\caption{Coherence}
	\label{fig_nshot_coh}
    \end{subfigure}
    \begin{subfigure}[b][][c]{.19\textwidth}
	\centering
        \includegraphics[width=\linewidth]{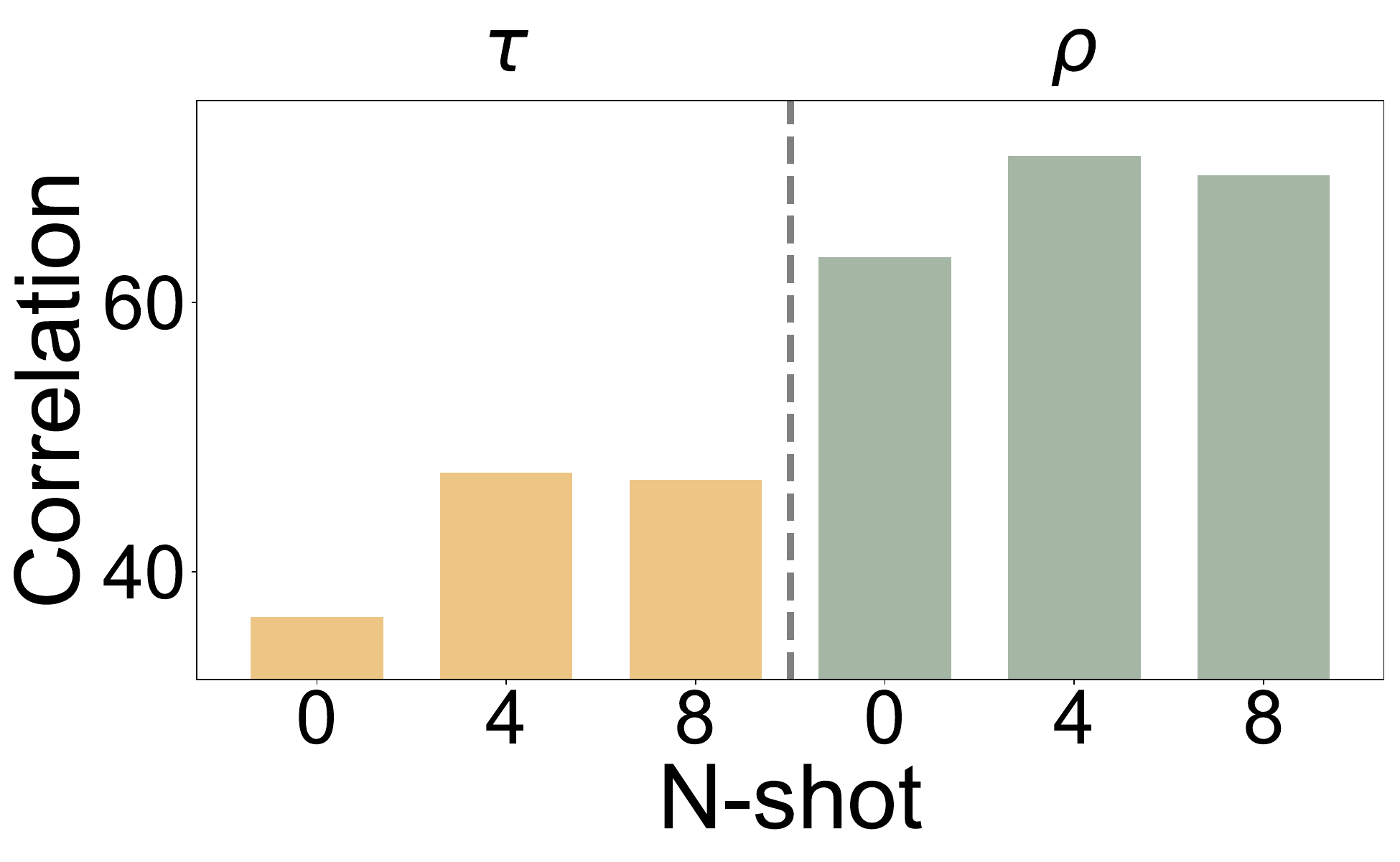}
	\caption{Consistency}
	\label{fig_nshot_con}
    \end{subfigure}
    \begin{subfigure}[b][][c]{.19\textwidth}
	\centering
        \includegraphics[width=\linewidth]{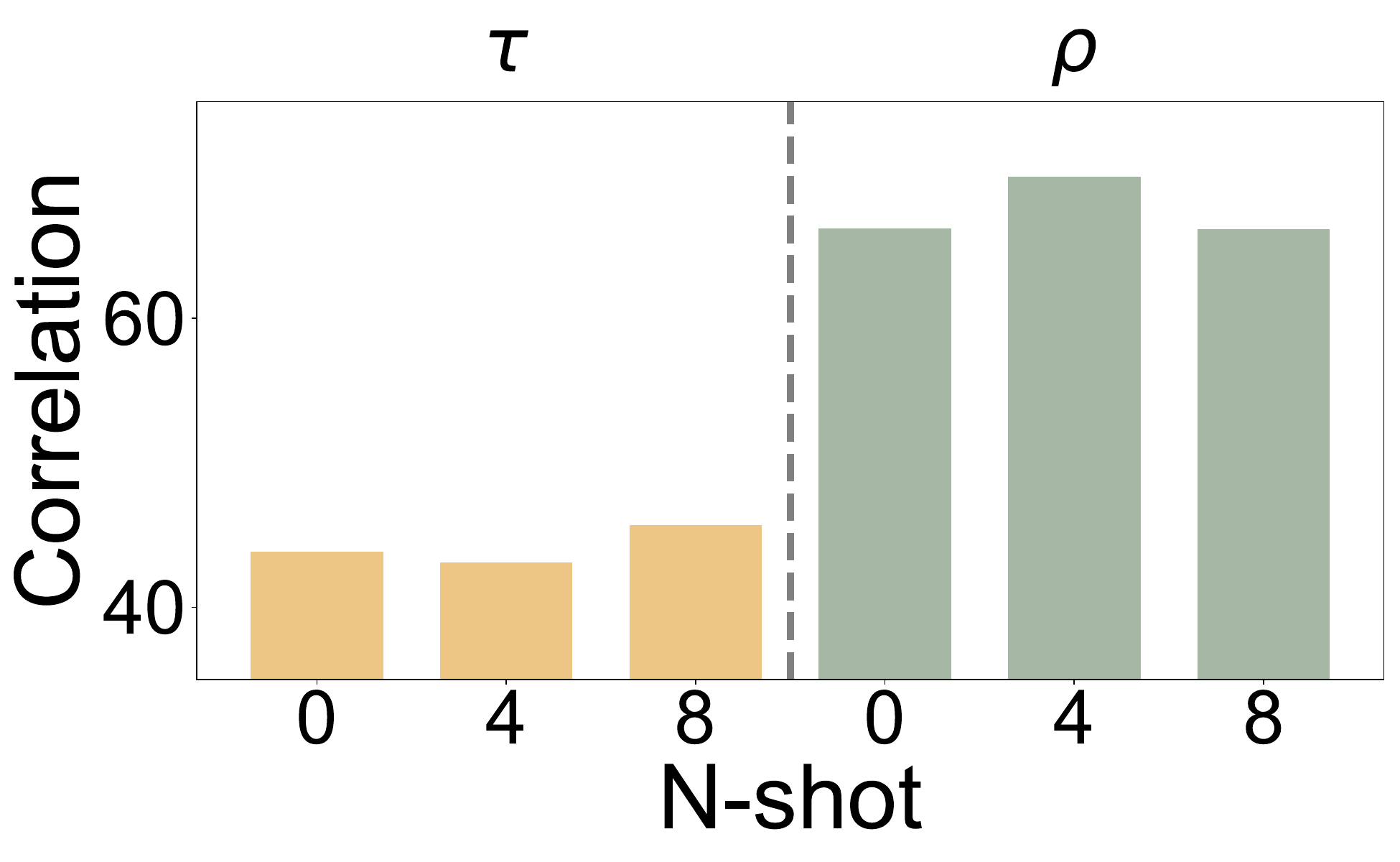}
	\caption{Fluency}
	\label{fig_nshot_flu}
    \end{subfigure}
    \begin{subfigure}[b][][c]{.19\textwidth}
	\centering
        \includegraphics[width=\linewidth]{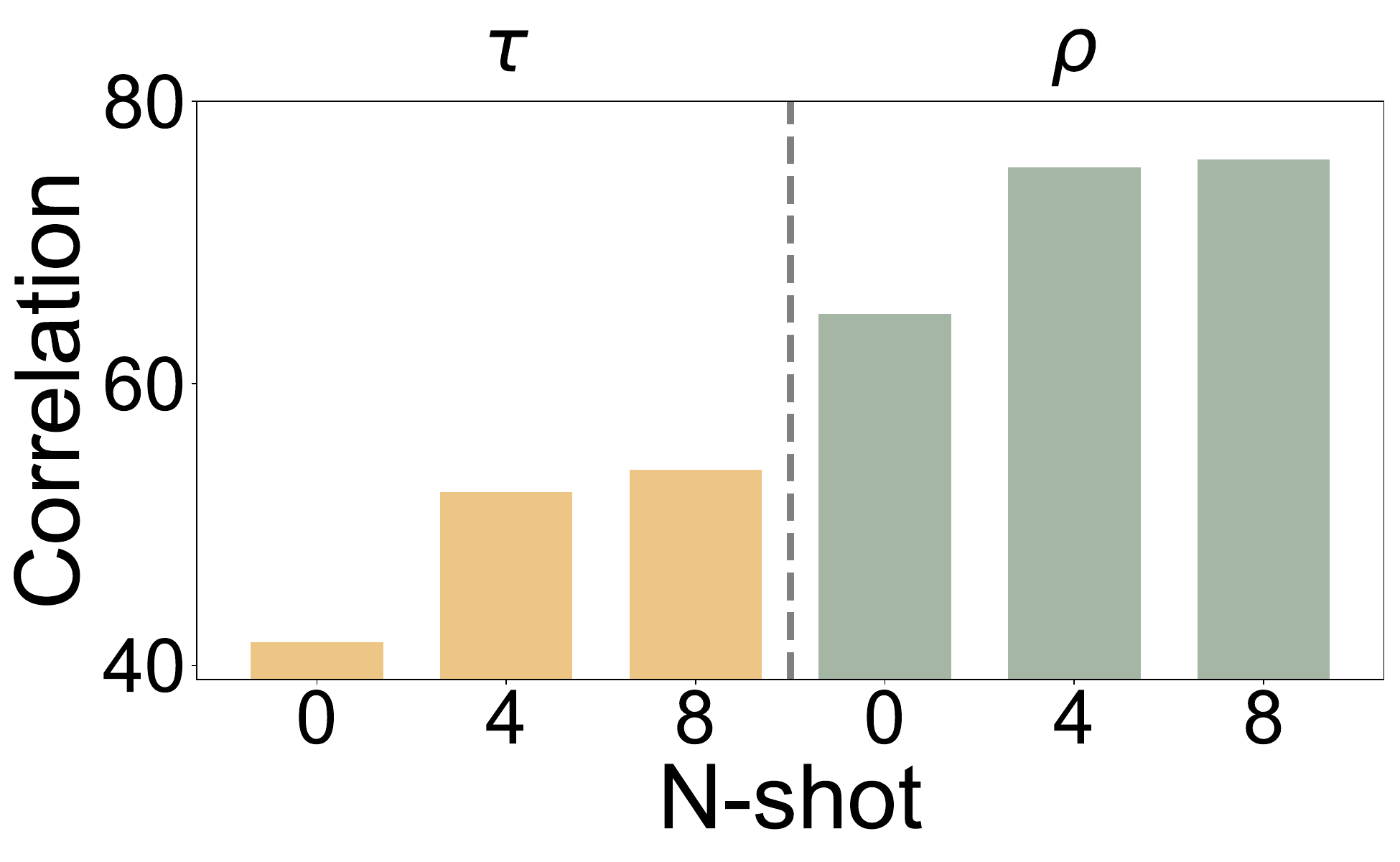}
	\caption{Relevance}
	\label{fig_nshot_rel}
    \end{subfigure}
    \begin{subfigure}[b][][c]{.19\textwidth}
	\centering
        \includegraphics[width=\linewidth]{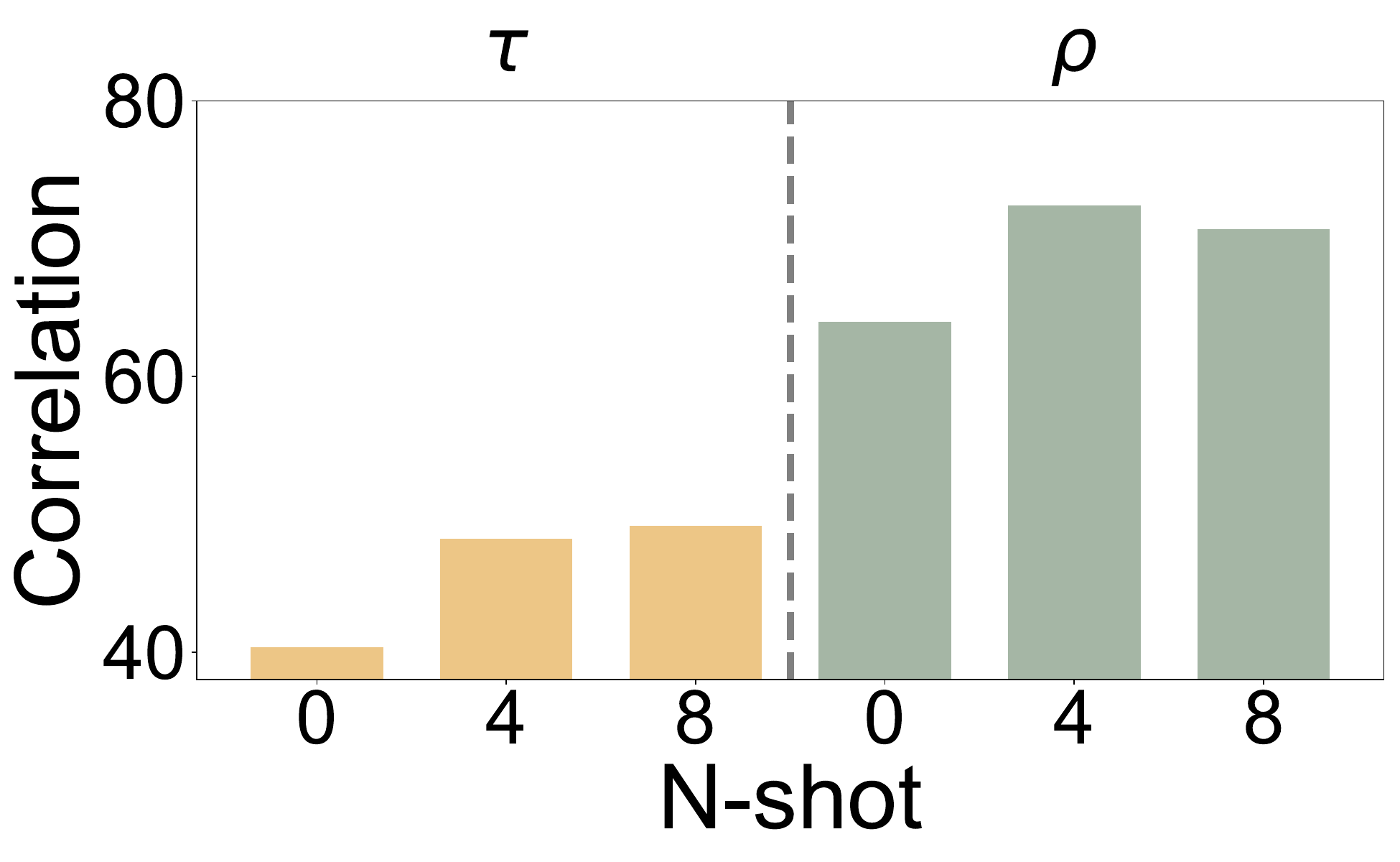}
	\caption{Average}
	\label{fig_nshot_avg}
    \end{subfigure}
\caption{Impact of demonstrations in LLM-based evaluator by \texttt{text-davinci-003} in the reference-free scenario.}
\label{figure_nshot}
\end{figure*}

\begin{figure*}[!t]
    \begin{subfigure}[b][][c]{.19\textwidth}
	\centering
        \includegraphics[width=\linewidth]{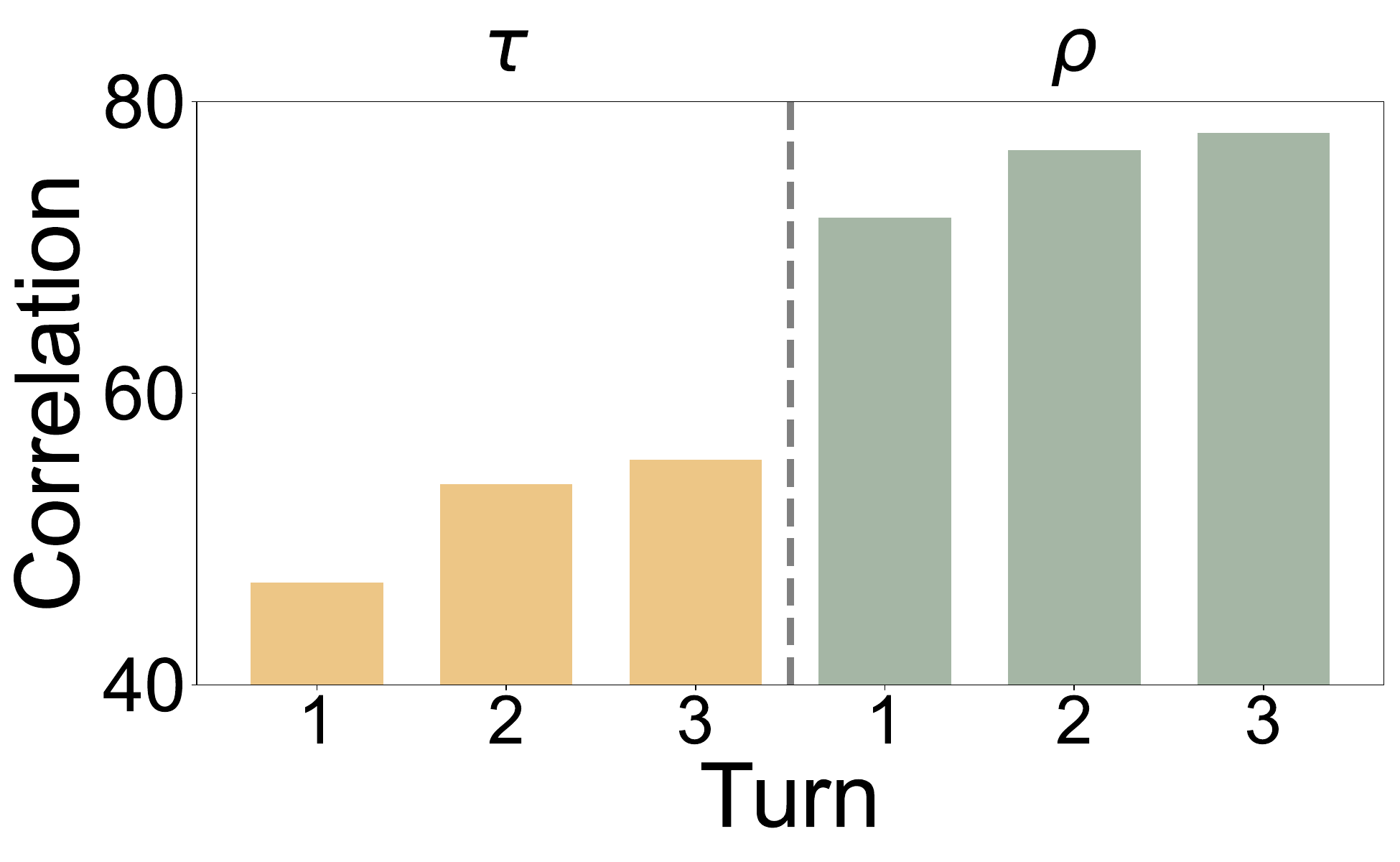}
	\caption{Coherence}
	\label{fig_turn_coh}
    \end{subfigure}
    \begin{subfigure}[b][][c]{.19\textwidth}
	\centering
        \includegraphics[width=\linewidth]{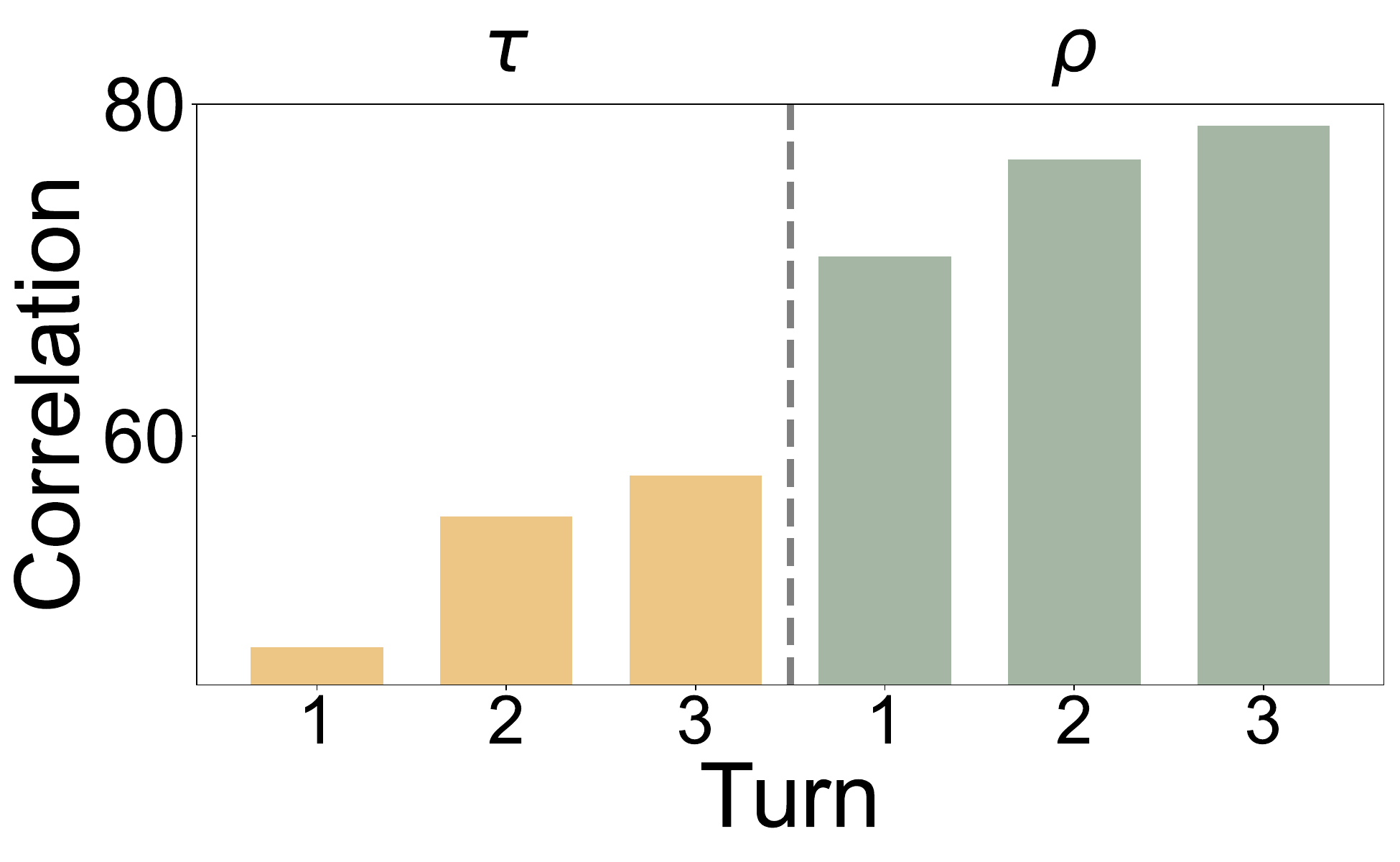}
	\caption{Consistency}
	\label{fig_turn_con}
    \end{subfigure}
    \begin{subfigure}[b][][c]{.19\textwidth}
	\centering
        \includegraphics[width=\linewidth]{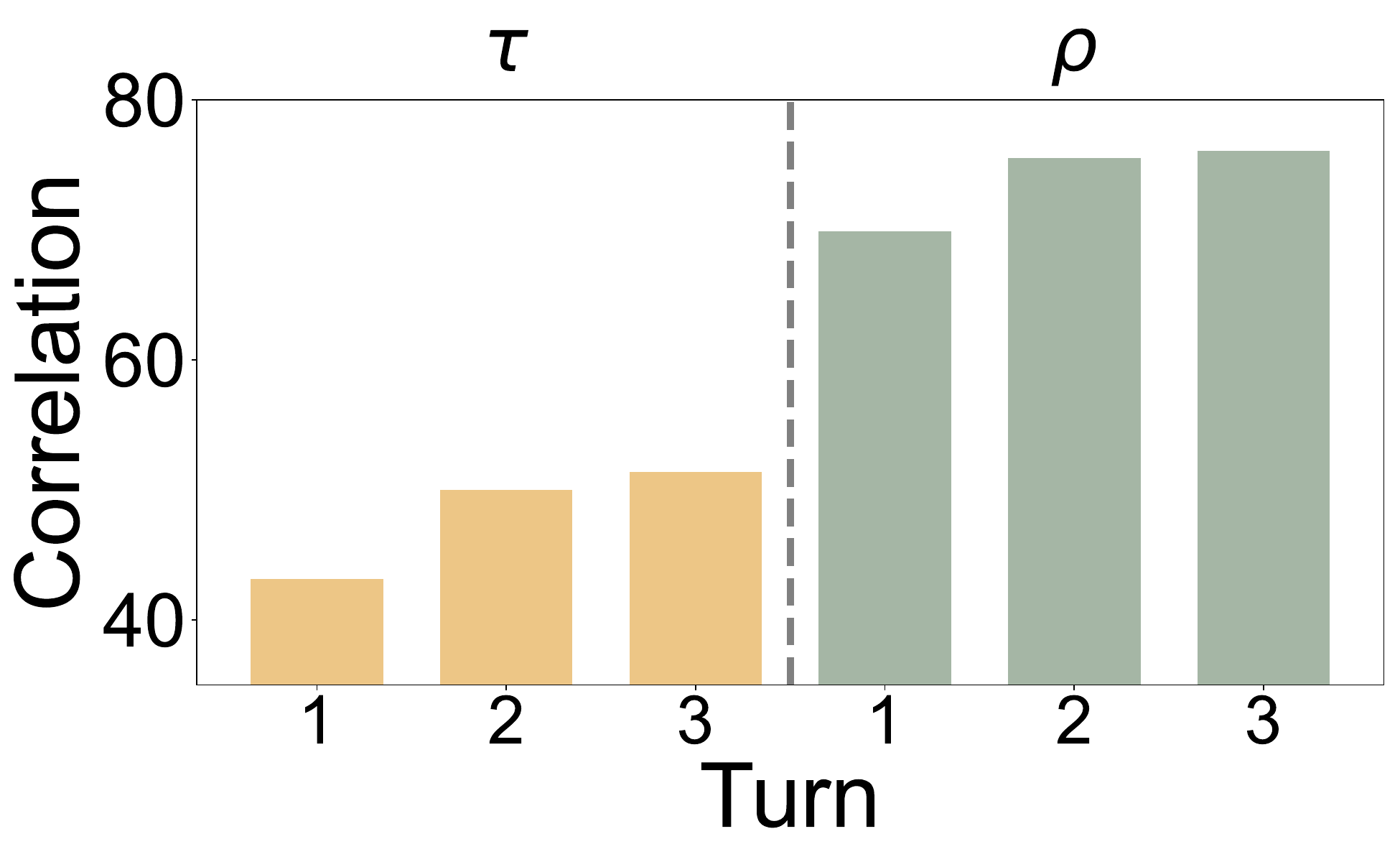}
	\caption{Fluency}
	\label{fig_turn_flu}
    \end{subfigure}
    \begin{subfigure}[b][][c]{.19\textwidth}
	\centering
        \includegraphics[width=\linewidth]{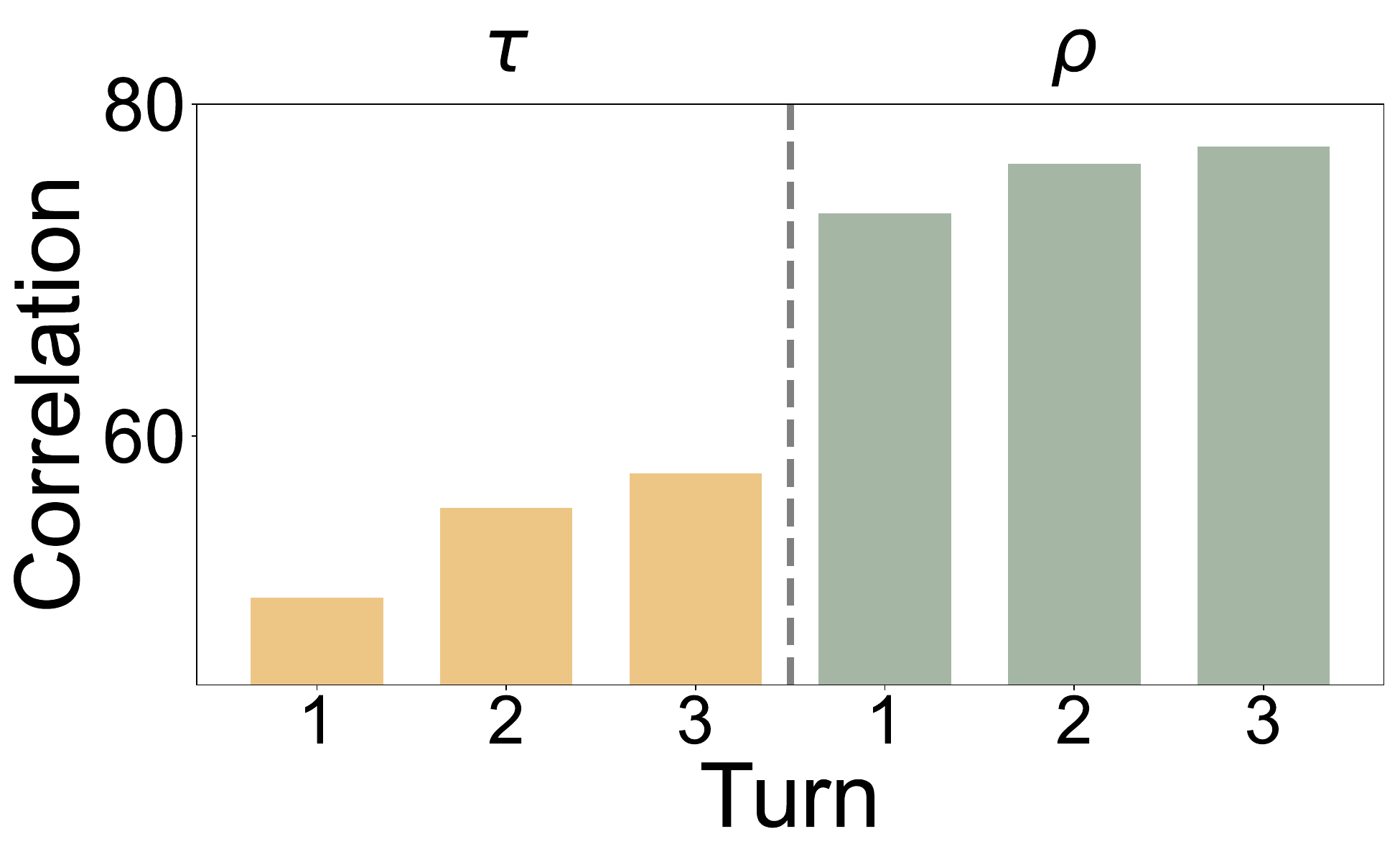}
	\caption{Relevance}
	\label{fig_turn_rel}
    \end{subfigure}
    \begin{subfigure}[b][][c]{.19\textwidth}
	\centering
        \includegraphics[width=\linewidth]{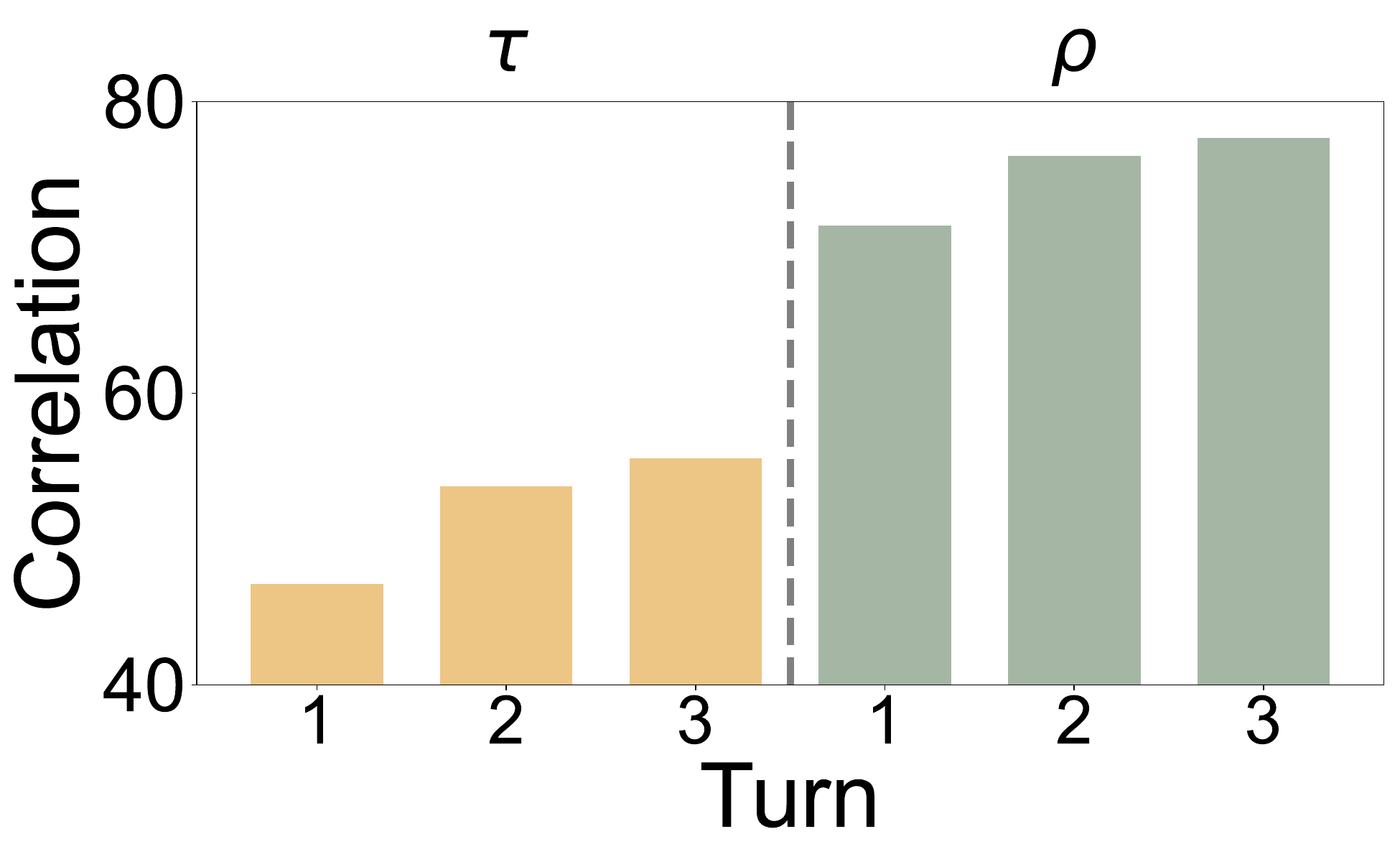}
	\caption{Average}
	\label{fig_turn_avg}
    \end{subfigure}
\caption{Impact of turns in LLM-based evaluator by \texttt{text-davinci-003} in the reference-free scenario.}
\label{figure_turn}
\end{figure*}

\subsection{Evaluation Criteria}
\label{criteria}
\subsubsection{Dimensions to Assess a Code Summary}
We evaluate generated code summaries across four key dimensions~\cite{fabbri2021summeval}:

\begin{itemize}[leftmargin=4mm, itemsep=0.05mm]
\item \textbf{Coherence (0-4).}
The summary should be logically organized, with a clear flow of ideas from sentence to sentence, forming a coherent description of the source code.

\item \textbf{Consistency (0-4).}
The summary must align with the \textit{facts} within the source code, e.g., specific statements, avoiding unsupported or hallucinated content.

\item \textbf{Fluency (0-4).}
The summary should be grammatically correct, well-structured, and free from repetition, formatting issues, and capitalization errors that impede readability.

\item \textbf{Relevance (0-4).}
The summary should capture the essential information from the source code, with penalties for redundancies and excessive details.
\end{itemize}
Following \cite{roy2021reassessing}, we engage 15 annotators—9 senior undergraduates and 6 graduate students, all with advanced English proficiency and 5+ years of software development experience—to ensure stable and reproducible evaluation scores.
Each annotator rates 300 samples on a 0-4 scale for coherence, consistency, fluency, and relevance of generated summaries based on the corresponding code snippets. 
We calculate Kendall’s Tau to verify agreement among the 15 annotators. Then, we average their scores for each generated summary.

\subsubsection{Metrics to Assess the Alignment with Human}
We calculate the correlation between automatic evaluation metrics with human scores 
employing Kendall-Tau correlation coefficient~\cite{kendall1945treatment} and Spearman correlation coefficient~\cite{ziegel2001standard}.

\mypara{Kendall-Tau Correlation Coefficient (\(\tau\))}
This metric evaluates the ordinal association between the datasets.
It provides a robust measure of the ordinal association between two measured quantities. For two datasets X and Y each with n data points, the Kendall-Tau correlation is defined as:
\begin{equation}
    \tau = \frac{2}{n(n-1)} \sum_{i<j} \text{sgn}(x_i - x_j) \cdot \text{sgn}(y_i - y_j)\,,
\end{equation}
where \( n \) is the number of pairs, and \( x_i, x_j, y_i, y_j \) are the ranks of the data points in the two datasets, X and Y, respectively.

\mypara{Spearman Correlation Coefficient (\(\rho\))}
It is particularly useful when the data does not conform to a normal distribution or when the relationship between variables is non-linear but monotonic. Spearman correlation \(\rho\) is given by:
\begin{equation}
    \rho = 1 - \frac{6 \sum_{i=1}^{n} d_i^2}{n(n^2 - 1)}\,,
\end{equation}
where \( d_i \) is the difference between the ranks of corresponding variables, and \( n \) is the total number of observations.

\subsection{RQ1: Performance of \system Evaluator}
\label{result_RQ1}
We investigate LLM-based evaluators by assigning diverse role-players, such as code reviewers, original code authors, code editors, and systems analysts, to assess code summarization across four dimensions: coherence, consistency, fluency, and relevance. 
We evaluate the performance of \system in its basic mode. 
This mode includes role profiles, task descriptions, evaluation criteria, demonstrations, and evaluation items. 
The default setting uses four demonstration examples.
It then proceeds with a single rating round in a score-only format.
We employ \texttt{gpt-4} as the backbone model and conduct experiments using the same dataset and summarization models—CodeNN, Deepcom, Astattgru, NCS, and Rencos—utilized in previous work~\cite{shi2022evaluation}. To ensure a fair comparison, we adhere to the methodology of prior work~\cite{shi2022evaluation} and evaluate 300 randomly sampled summaries generated by these models on the TL-CodeSum dataset.
Table~\ref{tab_overview} presents the results of experiments employing LLM-based evaluators across various role players, compared with conventional metrics. 
The results demonstrate a notable superiority of our proposed LLM-based evaluators in aligning with human assessments from different aspects (i.e., coherence, consistency, fluency, and relevance), significantly outperforming conventional metrics (i.e., BLEU, ROUGE, METEOR, and BERTScore).
Significantly, it is crucial to observe that these exceptional performances are achieved under both conditions, with and without the necessity of reference summaries, a requirement intrinsic to conventional metrics.
Notably, on average, our proposed \systemnospace, acting in the capacity of a code reviewer, surpasses the state-of-the-art BERTScore metric. This enhancement is evident in the increase of the Spearman correlation score from 64.32\% to 82.23\% with references, and to 81.59\% without references, affirming the efficacy of the LLM-based evaluator in the context of code summarization.
Additionally, we can see that reference-based evaluations outperform reference-free evaluations with minimal differences. This indicates that reference-free assessments are effective, as the scores are not strongly dependent on the reference, showing that LLMs can objectively evaluate code summaries under our framework.

\begin{tcolorbox}[left=1pt,right=1pt, top=1pt, bottom=0pt]
\textbf{Answer to RQ1.} 
Our proposed \system demonstrates a superior correlation with human scores across various dimensions, including coherence, consistency, fluency, and relevance, surpassing existing metrics.
\end{tcolorbox}

\begin{table*}[t!]
    \addtolength{\tabcolsep}{1.0pt}
    \centering
    \small
    \caption{The correlation of different prompting strategies on LLM-based evaluator backend by \texttt{text-davinci-003} with human evaluation. 
    }
    \setlength{\tabcolsep}{7pt} 
    \begin{tabular}{cc|cc|cc|cc|cc|cc}
        \hline
        \multicolumn{2}{c|}{\textbf{Ablations}}
        &\multicolumn{2}{c}{\textbf{Coherence}} & \multicolumn{2}{c}{\textbf{Consistency}} & \multicolumn{2}{c}{\textbf{Fluency}}&\multicolumn{2}{c}{\textbf{Relevance}} & \multicolumn{2}{c}{\textbf{Average}} \\
        \multicolumn{1}{c}{\textbf{CoT}} & \multicolumn{1}{c|}{\textbf{Forms}}
        &\multicolumn{1}{c}{\textbf{$\tau$}} & \multicolumn{1}{c}{\textbf{$\rho$}} 
        &\multicolumn{1}{c}{\textbf{$\tau$}} & \multicolumn{1}{c}{\textbf{$\rho$}} 
        &\multicolumn{1}{c}{\textbf{$\tau$}} & \multicolumn{1}{c}{\textbf{$\rho$}} 
        &\multicolumn{1}{c}{\textbf{$\tau$}} & \multicolumn{1}{c}{\textbf{$\rho$}} 
        &\multicolumn{1}{c}{\textbf{$\tau$}} & \multicolumn{1}{c}{\textbf{$\rho$}} \\
        \hline
        $\checkmark$	&Score only	&48.61\%&	72.48\%&	43.76\%&	67.89\%&	\textbf{44.11\%}&	\textbf{71.78\%}&	50.12\%&	71.44\%&	\cellcolor{gray!25}46.65\%&	\cellcolor{gray!25}70.89\%\\
        $\times$	&Score only	&50.08\%&	73.63\%&	47.39\%&	70.91\%&	43.13\%&	69.83\%&	52.31\%&	75.31\%&	\cellcolor{gray!25}48.23\%&	\cellcolor{gray!25}72.42\%\\
        $\times$	&Rate-explain	&41.37\%&	66.02\%&	46.13\%&	70.29\%&	38.18\%&	63.71\%&	45.95\%&	70.98\%&	\cellcolor{gray!25}42.90\%&	\cellcolor{gray!25}67.75\%\\
        $\times$	&Analyze-rate	&50.27\%&	71.2\%&	46.84\%&	69.46\%&	45.68\%&	66.16\%&	53.89\%&	75.9\%&	\cellcolor{gray!25}49.17\%&	\cellcolor{gray!25}70.68\%\\
        \hline
    \end{tabular}
    \label{tab_strategies}
\end{table*}

\subsection{RQ2: Influence of Evaluator Settings}
\label{result_RQ2}

\subsubsection{Influence of Evaluator Types}
We examine the impact of using different LLMs for evaluating code summarization, specifically the GPT-3.5 series models (i.e., \texttt{text-davinci-003} and \texttt{gpt-3.5-turbo}) and GPT-4 (i.e., \texttt{gpt-4}).
\texttt{text-davinci-003}\footnotemark[\value{footnote}]\footnotetext{
Note that as of January 2024, the \texttt{text-davinci-003} has been upgraded to \texttt{gpt-3.5-turbo-instruct}. More details are referred to OpenAI documentation: \url{https://platform.openai.com/docs/deprecations}.} provides high-quality outputs with reliable instruction-following ability, 
while \texttt{gpt-3.5-turbo} offers extended context length suitable for conversational applications.
\texttt{gpt-4} demonstrates high accuracy in complex problem-solving, making it effective for both interactive and traditional tasks.
Experimental results across various LLMs in the reference-free scenario are presented in Figure~\ref{figure_type}.
Obviously, the \texttt{gpt-4} model exhibits a superior performance over the \texttt{gpt-3.5} series of models. 
Moreover, the \texttt{text-davinci-003} demonstrates well-rounded performance across overall evaluations. 
While \texttt{gpt-4} outperforms the \texttt{gpt-3.5} series, its higher API cost should be considered. 
Thus, we recommend choosing different LLMs as evaluators based on performance needs and budget considerations.

\subsubsection{Influence of Number of Demonstration Examples} To analyze the impact of demonstration examples on our proposed LLM-based evaluator for code summarization, we target the \texttt{text-davinci-003} model and vary the number of demonstration examples used: 0, 4, and 8. 
Figure~\ref{figure_nshot} presents the results for varying numbers of demonstration examples. 
Notably, using 4 demonstration examples generally yields superior performance, with an average improvement of 3.59\% in Spearman correlation compared to evaluations without demonstration examples. 
Thus, we recommend this configuration for enhanced evaluation performance.

\subsubsection{Influence of Turns} We examine rating generability and robustness by varying the number of turns and averaging scores across rounds. Specifically, we explore the impact of turns on \texttt{text-davinci-003} evaluators when assessing code summarization by varying the number of turns from 1 to 3. 
From Figure~\ref{figure_turn}, it is evident that a higher number of turns generally correlates with improved performance for LLM-based evaluators.
Specifically, the average scores derived from three turns approximate human scores, with a Kendall-Tau correlation of 55.54\% and a Spearman correlation of 77.51\%.

\subsubsection{Influence of Prompt Settings}
We conduct an ablation study to analyze the impact of \textit{evaluation steps prompt} and \textit{rating forms prompt}, as introduced in Sec.~\ref{strategies}. 
Table~\ref{tab_strategies} shows the correlation between different prompting strategies with human evaluation results. 
From this table, it is evident that the guidance provided by CoT in the evaluation steps is not universally beneficial. The findings indicate its effectiveness primarily in appraising fluency, showcasing a noteworthy improvement of 0.98\% in Kendall-Tau correlation (\(\tau\)) and 1.95\% in Spearman correlation (\(\rho\)).
Interestingly, we find the \textit{analyze-rate form} consistently outperforms the \textit{rate-explain form} across all aspects. 
It is important to highlight that employing these form-based strategies does not necessarily result in improved performance when compared to a simplistic \textit{score only} approach, where LLMs are simply tasked with providing a numerical score.

\begin{tcolorbox}[left=1pt,right=1pt, top=1pt, bottom=1pt]
\textbf{Answer to RQ2.} 
While \system surpasses traditional metrics in achieving best performance for code summarization evaluation, it requires careful prompt design and the selection of an appropriate base LLM.
\end{tcolorbox}

\subsection{RQ3: Re-Evaluation of Current Models}\label{result_RQ3}
\begin{table}[t!]
    \centering
    \caption{
    Reevaluation of code summarization models with \system (with \texttt{text-davinci-003}), normalizing the average scores to a range of 0\% to 100\%.
    }
    \setlength{\tabcolsep}{7pt} 

    \begin{tabular}{l|cccc}
        \hline
        \textbf{Model} &\multicolumn{1}{c}{\textbf{Coherence}} & \multicolumn{1}{c}{\textbf{Consistency}} & \multicolumn{1}{c}{\textbf{Fluency}}&\multicolumn{1}{c}{\textbf{Relevance}}  \\
        \hline
        CodeNN	& 24.00\%&	22.50\%&	29.50\%&	27.50\% \\
        Deepcom	&	16.50\%&	21.25\%&	18.25\%& 18.25\%\\
        Astattgru	&	36.25\%&	28.25\%&	49.75\%&	38.00\%\\
        Rencos	&	54.00\%&	48.50\%&	66.25\%&	51.50\%\\
        NCS	&	56.25\%&	54.75\%&	69.50\%&	58.25\%\\
        ChatGPT	&	\textbf{89.50}\%&	\textbf{95.00}\%&	\textbf{91.75}\%&	\textbf{90.75}\%\\
        \hline
    \end{tabular}
    \label{tab_eval}
\end{table}
\begin{figure*}[!t]
\centering
\includegraphics[width=0.98\textwidth]
{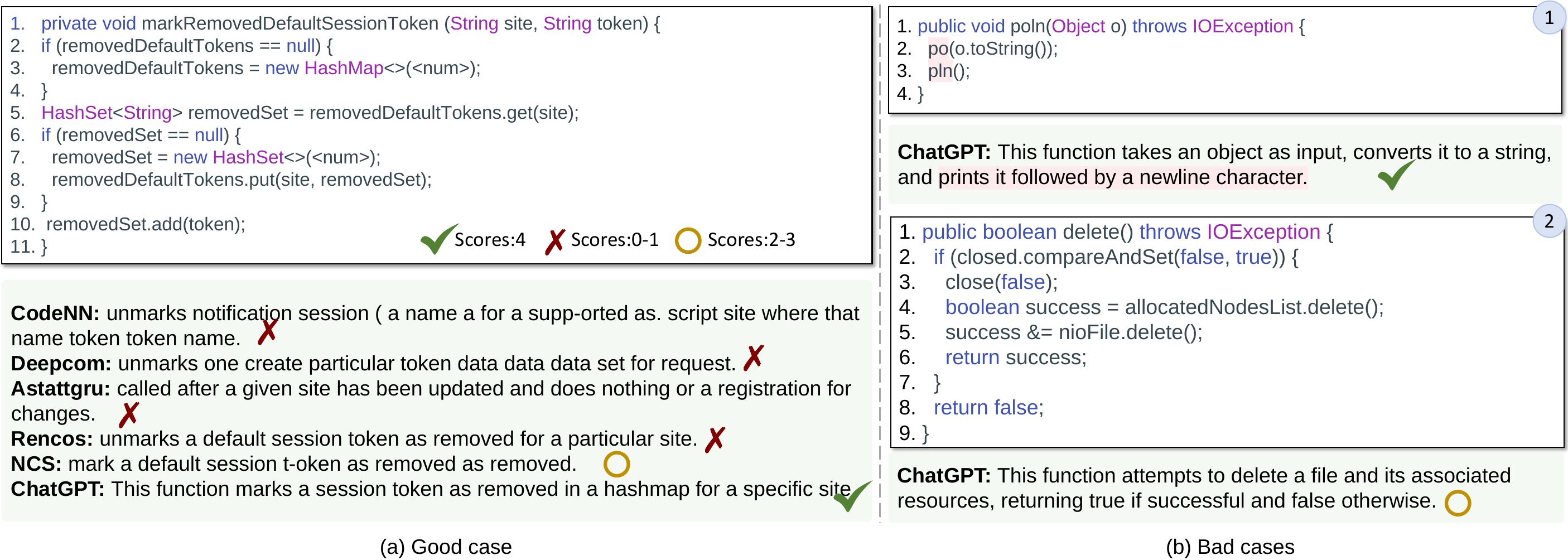}
\caption{Case studies to investigate when 
\system
works and fails. 
}
\label{fig_casestudy}
\end{figure*}
We systematically re-evaluate established code summarization models, e.g., CodeNN, Deepcom, Astattgru, Rencos, NCS, and ChatGPT, employing LLM-based evaluators \system equipped by \texttt{text-davinci-003}. Our analysis utilizes the TL-CodeSum dataset, from which we randomly choose 100 summaries for evaluation.

Table~\ref{tab_eval} shows the evaluation results.
From this table, we can observe a notable superiority of ChatGPT across various dimensions, including coherence, consistency, fluency, and relevance, as assessed by our newly proposed LLM-based evaluators, when compared to conventional baselines.
In particular, code summaries generated by ChatGPT exhibit a remarkable achievement of approximately 90\% across all four specified dimensions, surpassing significantly other neural models designed for code summarization. For instance, concerning coherence, ChatGPT demonstrates notable improvement over the state-of-the-art NCS model, elevating performance from 56.25\% to 89.50\%.
Interestingly, this finding contrasts with a prior research work~\cite{sun2023automatic}, which indicated ChatGPT's lower performance compared to specialized code summarization models (e.g. NCS) using BLEU, METEOR, and ROUGE-L metrics. 
Our research demonstrates ChatGPT's effectiveness through LLM evaluators, 
emphasizing the importance of developing robust evaluation methods for code summarization performance assessment.

\begin{tcolorbox}[left=1pt,right=1pt, top=1pt, bottom=1pt]
\textbf{Answer to RQ3.} 
Our proposed \system shows that ChatGPT has superior performance in code summarization than other neural models, exhibiting a closer alignment with human judgments.
\end{tcolorbox}

\section{Discussion}
\subsection{When do LLM-based Evaluators Work and Fail?}
To enhance our comprehension of the conditions under which LLM-based evaluators succeed or encounter challenges, we conduct an exhaustive case study, encompassing both a good case and a case characterized by errors.
In Figure~\ref{fig_casestudy} (a), a good case is presented, wherein both a code snippet and its corresponding summaries generated by various neural models are provided. In analyzing this instance, it becomes apparent that the code summary produced by ChatGPT exhibits notable differences compared to other summaries. However, employing our proposed LLM-based evaluator reveals that the code summary generated by ChatGPT attains the highest score of 4, signifying superior quality.
Upon manual inspection of the quality of generated summaries, we acknowledge that the summaries produced by ChatGPT exhibit superior quality compared to those generated by other models, showcasing a remarkable alignment with human judgments. This superiority can be attributed to the impressive summarization capabilities inherent in LLMs.

Moreover, we present two scenarios to illustrate instances where LLM-based evaluators may fail to assess code summarization models, as depicted in Figure~\ref{fig_casestudy} (b).
In the first error case, the LLM evaluator gives the consistency of the summary a score of 4. However, from our manual inspection, we can see a hallucination content ``\textit{prints it followed by a newline character}'', which is extended from the words ``\texttt{po}'' and ``\texttt{pln}''. The LLM-based evaluator is unaware of the fact that ``\texttt{po}'' and ``\texttt{pln}'' functions are defined outside the scope of the provided code. This oversight might originate from pre-training biases, where LLMs prematurely learned to associate these abbreviations with common methods.
In the second error case, 
we can see that the code summary generated by ChatGPT receives a score of 2, despite being deemed concise and effective by human annotators. This could be due to the model's subjective interpretation of an ideal summary, leading it to perceive the provided summary as lacking in detail.

It is important to recognize that similar subjective preferences could be held by human evaluators as well, influencing their judgments and introducing variability into the evaluation process. 
Our work advocates for a balanced approach, where LLM evaluation serves as a complementary tool rather than a replacement for human evaluation. Both human and LLM evaluation have their own strengths and weaknesses and can be effectively combined.
We encourage future researchers and practitioners in the field of code summarization to consider the dual use of LLM and human evaluations.

\subsection{Implications}
In this study, we have obtained several significant implications that offer valuable insights for further study.

\mypara{Implication 1}
Our comprehensive investigations demonstrate that LLMs can indeed serve as effective evaluators for code summarization, 
surpassing established evaluation metrics (i.e., BLEU-R, ROUGE-L, METEOR, and BERTScore) in their alignment with human judgments.
This can inspire our research community to develop enhanced LLMs for assessing the efficacy of code intelligence tasks, encompassing code generation, vulnerability detection, and many others.

\mypara{Implication 2}
Our comprehensive ablation studies underscore the importance of meticulous attention to designing effective prompting strategies, careful selection of the LLM as the backbone, and thoughtfully setting the roles for each LLM agent.
This can provide valuable insights for the research community to enhance the design of LLMs, thereby improving their effectiveness in evaluating code intelligence tasks.

\mypara{Implication 3}
Our LLM-based evaluators reveal that ChatGPT outperforms other neural models in code summarization, providing strong evidence for the efficacy of LLMs in understanding code and generating precise summaries.

\subsection{Threats to Validity}
\mypara{Limited Roles and Agents}
In this paper, we assign distinct roles to an LLM agent, encompassing functions as code reviewers, original code authors, code editors, and systems analysts in the design of prompt strategies. Additional player roles, including software tester and software project manager, merit further exploration. The investigation of these roles is deferred to our future work.
Furthermore, in this paper, we initially utilize a single LLM agent to act in multiple roles. However, recognizing the importance of multi-agent collaboration, we encourage a multi-agent setting, where each LLM agent serves an individual role.  
Within this framework, an essential focus needs to be placed on investigating collaboration and communication dynamics among multiple agents.

\mypara{Prompt Engineering}
As described in Section \ref{sec_prompt_stragety}, our LLM-based evaluators depend on the prompts we craft. Typically, these prompts are manually designed, necessitating substantial human effort.
This will pose a challenge to the broad applicability of our LLM-based evaluators for diverse code intelligence tasks, limiting their potential for effective generalization.

\mypara{Limited Ground-Truth Summary}
Our evaluation relies significantly on ground-truth code summaries, which we obtain through human annotation. In this study, we adopt prior work~\cite{shi2022evaluation} amassing a dataset comprising 
300 code summaries. 
We treat these human annotations as the ground truth for our evaluation and recognize the need to expand the dataset in both size and diversity of code types.
Note that we do not consider the LLMs' prior exposure to these datasets as a source of data leakage or bias, since the model’s evaluation focuses on understanding the code and assessing its alignment with the summaries, rather than generating ground-truth-like summaries.
As part of our future work, we intend to expand our assessor pool and collect additional ground-truth summaries to enhance the comprehensiveness of our evaluation.

\section{Related Work}

\subsection{Code Summarization}
Source code summarization plays a critical role in improving program comprehension and maintenance.
This task, traditionally labor-intensive and time-consuming, often results in descriptions that are incomplete, incorrect, or outdated~\cite{briand2003software,forward2002relevance,tilley1992documenting}. 
Recently, deep learning-based techniques have driven the development of automated approaches to code summarization~\cite{wang2020reinforcement,guo2022modeling,wan2018improving,10.1145/3664597, Zhangsurvey}, such as DeepCom~\cite{hu2018deep}, NCS~\cite{ahmad2020transformer}, and SIT~\cite{wu2020code}, utilizing large-scale code-summaries corpora for training generative models to translate code into natural language summaries~\cite{alon2018code2seq,gros2020code,iyer2016summarizing,leclair2019neural,shi2022evaluation}. 
Furthermore, pre-trained models have been adapted to enhance code summarization, as evidenced in works like CodeBERT~\cite{feng2020codebert} and CodeT5~\cite{wang2021codet5}. 
More recently, the emergence of LLMs has garnered significant interest, prompting numerous studies to investigate its potential for code summarization~\cite{sun2023automatic,khan2022automatic,ahmed2024automatic, cai2024fly,su2024distilled,sun2024source}.
For example, Su et al.~\cite{sun2024source} investigated diverse prompting techniques, including zero-shot, few-shot, chain-of-thought, critique, and expert methods, to adapt LLMs for code summarization.
Complementary to these studies, this paper focuses on evaluating the quality of generated code summaries.

\subsection{Code Summarization Evaluation} 
Evaluating code summarization is challenging due to the inherently open-ended nature of the task.
Existing evaluation approaches can be generally divided into two categories: automatic and human evaluations.
Automatic methods typically rely on metrics such as BLEU, METEOR, and ROUGE-L to compare generated summaries against reference summaries. 
When reference summaries are unavailable, human-generated summaries are sometimes used as benchmarks. An in-depth analysis of these metrics, particularly BLEU, investigated their correlation with human perception~\cite{shi2022evaluation}. 
However, as Stapleton et al.~\cite{stapleton2020human} argued, these metrics often focused more on syntactic rather than semantic aspects, and did not fully capture the impact of machine-generated code summaries on human comprehension or productivity. Moreover, this gap is evident in the work that highlights the limitations of these metrics in evaluating the creative and diverse outputs from models like ChatGPT~\cite{khan2022automatic}. On the other hand, human evaluations, which entail participants' assessment of the summarization quality, can mitigate some of these shortcomings. However, these evaluations are often labor-intensive and infrequently conducted in practice.

\subsection{Natural Language Generation Evaluators}
Recently, various works have been proposed to evaluate the capabilities of LLMs in generative tasks~\cite{chiang2023can,fu2023gptscore,wang2023chatgpt,chenmllm}. 
DRPE~\cite{wu2023large} introduced a roleplayer-based prompting strategy, enabling LLMs to evaluate generated text against golden references with human-like proficiency. Contemporary research advocates for using LLMs as reference-free evaluators. For instance, GEMBA~\cite{kocmi2023large} demonstrated the state-of-the-art capability of GPT-based translation quality assessment. Wang et al.~\cite{wang2023chatgpt} conducted a meta-evaluation on ChatGPT, showcasing its reliability in various task-specific and aspect-specific evaluations, correlating closely with human judgment. Similarly, G-EVAL~\cite{liu2303g} employed LLMs with CoT to achieve higher human correspondence in tasks like text summarization and dialogue generation. Further, Chiang and Lee~\cite{chiang2023can} explored LLMs as alternatives to human evaluators, with subsequent analysis in~\cite{chiang2023closer} offering guidelines for using ChatGPT as an automatic evaluation tool. 
Moreover, Chan et al.~\cite{chan2023chateval} proposed a framework involving multiple evaluator agents to simulate the process of reaching consensus among human annotators.
Furthermore, the exploration of LLMs has extended into code-related tasks. ICE-Score~\cite{zhuo2024ice} leveraged LLMs to assess the quality of code without the need for oracles or references, setting a new benchmark for the evaluation of code generation.
To assess these LLM-based evaluators, Zeng et al. 
\cite{zeng2023evaluating} introduced a robust meta-evaluation benchmark for selecting evaluators knowledgeably, while Doostmohammadi et al. \cite{doostmohammadi2024reliable} delved into the reliability of automatic evaluation methods. 
Furthermore, Shankar et al.~\cite{10.1145/3654777.3676450} presented a mixed-initiative approach to validate LLM-assisted evaluations, demonstrating that LLM evaluation functions can indeed be aligned and validated against human preferences, thus reinforcing the feasibility of reliable LLM evaluation.

\section{Conclusion}
In this paper, we have explored  
the capability of LLMs 
to evaluate code summarization models. We propose \systemnospace, a novel LLM-based metric for assessing the quality of code summaries in four dimensions: coherence, consistency, fluency, and relevance. Particularly, we focus on the role-playing ability of LLMs, simulating various personas such as code reviewers, original code authors, code editors, and systems analysts. 
Moreover, we investigate the effects of prompt strategies and assess the robustness of the metrics. Experimental results reveal that our approach aligns more closely with human evaluations, presenting a promising alternative to traditional metrics.

\mypara{Data Availability}
All the source code and experimental data referenced in this paper can be accessed at:
\texttt{\url{https://github.com/CGCL-codes/naturalcc/tree/main/examples/CodeSum-Eval}}~\cite{wan2022naturalcc}.

\balance

\bibliographystyle{IEEEtran}
\bibliography{IEEEabrv,ref}

\end{document}

%% file: preamble.tex
\usepackage{cite}
\usepackage{amsmath,amssymb}
\usepackage{amsfonts}
\usepackage{algorithmic}
\usepackage{graphicx}
\usepackage{textcomp}
\usepackage{xcolor}
\def\BibTeX{{\rm B\kern-.05em{\sc i\kern-.025em b}\kern-.08em
    T\kern-.1667em\lower.7ex\hbox{E}\kern-.125emX}}

\usepackage{longfbox}
\usepackage[american]{babel}
\usepackage{microtype}
\usepackage{subcaption}
\usepackage[ruled,,linesnumbered ]{algorithm2e}
\usepackage[most]{tcolorbox}
\usepackage{balance} 
\usepackage{enumitem} 
\usepackage{color}
\usepackage{url}
\usepackage{arydshln}
\usepackage{bm}
\usepackage{multirow}

\usepackage{bbm}
\usepackage{makecell}
\usepackage{booktabs}
\usepackage{amsthm}

\newtheoremstyle{bolddefinition} 
{}          
{}          
{}          
{}          
{\bfseries} 
{.}         
{.5em}         
{\thmname{#1}\thmnumber{ #2}\thmnote{ (\bfseries #3)}}          

\theoremstyle{bolddefinition}

\usepackage{tcolorbox}
\tcbuselibrary{skins, breakable, theorems}
\usepackage{colortbl}
\newcommand{\system}{\textsc{CodeRPE \xspace}}%
\newcommand{\systemnospace}{\textsc{CodeRPE}}%

\newcommand{\mypara}[1]{\smallskip \noindent\textbf{#1.} \xspace}

\newcount\DraftStatus  
\definecolor{darkgreen}{rgb}{0,0.5,0} 
\definecolor{purple}{rgb}{1,0,1} 
\definecolor{todocolor}{rgb}{0.9,0.1,0.1} 
\definecolor{fixcolor}{rgb}{0.1,0.7,0.3} 
\definecolor{wycolor}{rgb}{0.9,0.1,0.1} 
\definecolor{hycolor}{rgb}{0.7,0.7,0.3}
\definecolor{wtzcolor}{rgb}{0.9,0.6,0.8} 
\definecolor{yecolor}{rgb}{0.85,0.65,0.2} 
\definecolor{wyangcolor}{rgb}{0.65, 0.902, 0.2}
\definecolor{ashgrey}{rgb}{0.7, 0.75, 0.71}
\definecolor{grey}{rgb}{0.6,0.6,0.6}

\newcommand{\nbc}[3]{\ifnum\DraftStatus=1
	{\colorbox{#3}{\bfseries\sffamily\scriptsize\textcolor{white}{#1}}}
	{\textcolor{#3}{\sf\small$\blacktriangleright$\emph{#2}$\blacktriangleleft$}}
\fi}

\newcommand{\draftnote}[2]{\ifnum\DraftStatus=1
	\marginpar{
		\tiny\raggedright
		\hbadness=10000
		\def\baselinestretch{0.8}
		\textcolor{#1}{\textsf{\hspace{0pt}#2}}}
\fi}